\documentclass[11pt,reqno]{amsart}
\usepackage{geometry}
\usepackage{enumerate}
\usepackage{latexsym,booktabs}
\usepackage{amsmath,amssymb}
\usepackage{graphicx}
\usepackage{color}
\usepackage{amsmath}
\usepackage{scalefnt}
\usepackage{setspace}
\usepackage[colorlinks=false,hidelinks]{hyperref}
\usepackage{amsthm}
\usepackage{mathrsfs}
\usepackage{multirow}
\usepackage{fancyvrb}
\usepackage{comment}
\usepackage{framed}
\usepackage{multicol}
\usepackage{multirow}
\usepackage{eurosym}
\usepackage{amsopn}
\usepackage{amssymb}
\usepackage[dvipsnames]{xcolor}
\usepackage{array}
\usepackage{amsthm}
\usepackage[toc,page]{appendix}
\usepackage{wrapfig}
\usepackage{subcaption}
\usepackage{indentfirst}
\usepackage{physics}
\usepackage{rotating}
\usepackage{soul}
\usepackage{color}
\usepackage{capt-of}
\usepackage{listings}
\usepackage{natbib}
\usepackage{pdflscape}
\usepackage{mathtools}
\usepackage[foot]{amsaddr}
\usepackage[font=small,labelfont=bf]{caption}
\usepackage{lineno}
\usepackage[ruled,vlined]{algorithm2e}

\numberwithin{Theorem}{section}
\numberwithin{Corollary}{section}

\numberwithin{Proposition}{section}
\numberwithin{Definition}{section}
\numberwithin{Lemma}{section}
\numberwithin{equation}{section}

\newcommand*{\maxam}{\text{maxAM}}
\newcommand*{\am}{\text{AM}}
\newcommand{\mybar}[1]{\makebox[0pt]{$\phantom{#1}\overline{\phantom{#1}}$}#1}

\newcommand{\N}{\mathcal{N}}
\newcommand{\bN}{\mybar{\mathcal{N}}}
\newcommand{\D}{\mathcal{D}}
\newcommand{\bD}{\mybar{\mathcal{D}}}
\newcommand{\HH}{\mathcal{H}}

\title[K\lowercase{w}ARG]{\vspace{-20pt}K\lowercase{w}ARG: Parsimonious reconstruction of ancestral \\ \vspace{4pt} recombination graphs with recurrent mutation\vspace{-1pt}}
\date{May 6, 2021}
\author{Anastasia Ignatieva$^1$}
\address{$^1$ \textnormal{Department of Statistics, University of Warwick, Coventry CV4 7AL, UK}}
\email{anastasia.ignatieva@warwick.ac.uk}

\author{Rune B. Lyngs\o\,$^2$}
\address{$^2$ \textnormal{Department of Statistics, University of Oxford, 24-29 St Giles', Oxford OX1 3LB, UK}}

\author{Paul A. Jenkins\,$^1$\,$^3$\,$^4$}
\address{$^3$ \textnormal{Department of Computer Science, University of Warwick, Coventry CV4 7AL, UK}}

\author{Jotun Hein\,$^2$\,$^4$}
\address{$^4$ \textnormal{The Alan Turing Institute, British Library, London NW1 2DB, UK}}

\makeatletter
\renewcommand\@setemails{%
\mbox{{\itshape E-mail}:\space}{\ttfamily\emails}.
}
\makeatother

\begin{document}

\captionsetup{width=0.9\textwidth}
\vspace*{-5pt}

\maketitle

\vspace*{-20pt}
\begin{abstract}
The reconstruction of possible histories given a sample of genetic data in the presence of recombination and recurrent mutation is a challenging problem, but can provide key insights into the evolution of a population. We present KwARG, which implements a parsimony-based greedy heuristic algorithm for finding plausible genealogical histories (ancestral recombination graphs) that are minimal or near-minimal in the number of posited recombination and mutation events. Given an input dataset of aligned sequences, KwARG outputs a list of possible candidate solutions, each comprising a list of mutation and recombination events that could have generated the dataset; the relative proportion of recombinations and recurrent mutations in a solution can be controlled via specifying a set of `cost' parameters. We demonstrate that the algorithm performs well when compared against existing methods. The software is made available on GitHub.

\end{abstract}



\section{Introduction}

For many species, the evolution of genetic variation within a population is driven by the processes of mutation and recombination in addition to genetic drift. A typical mutation affects the genome at a single position, and may or may not spread through subsequent generations by inheritance. Recombination, on the other hand, occurs when a new haplotype is created as a mixture of genetic material from two different sources, which can drive evolution at a much faster rate. The detection of recombination is an important problem which can provide crucial scientific insights, for instance in understanding the potential for rapid changes in pathogenic properties within viral populations \citep{viruses_recombine}.

\begin{figure}[]
\centering
        \includegraphics[width=0.85\linewidth]{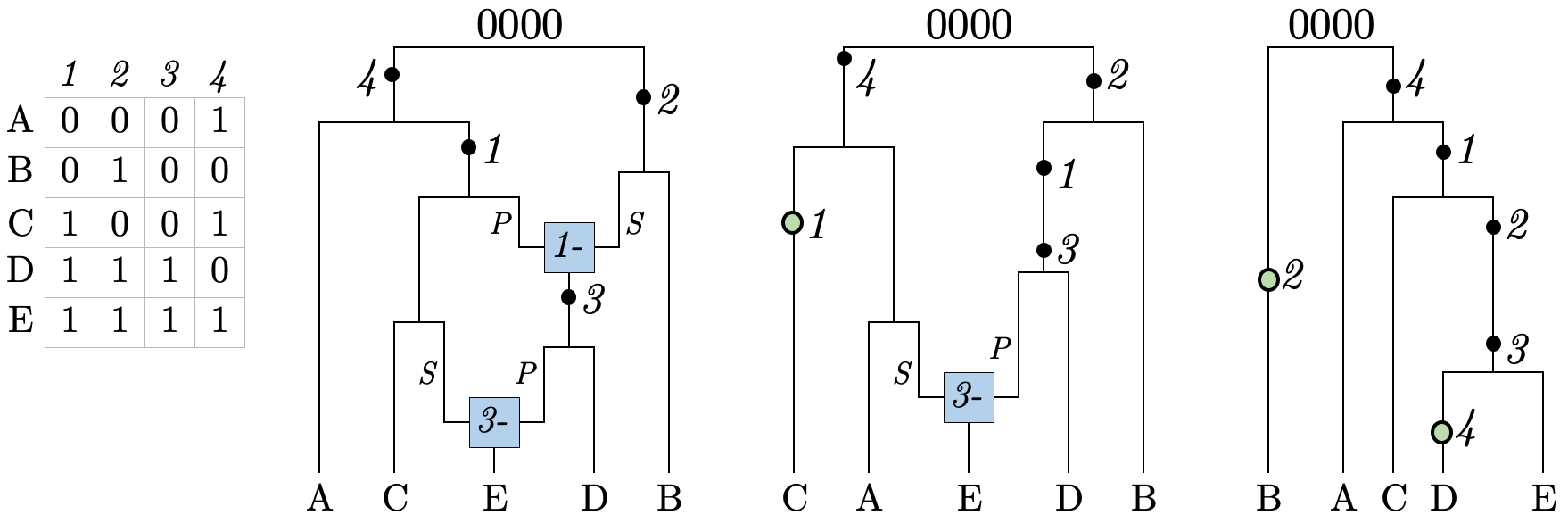}
         \setlength{\belowcaptionskip}{0pt}
    \caption{Three examples of ARGs. The dataset is shown on the left in binary format, with 0's and 1's corresponding to the ancestral and mutant state at each site, respectively. Mutation events are shown as black dots and labelled by the site they affect; green filled circle corresponds to a recurrent mutation. Recombination nodes (in blue) are labelled with the recombination breakpoint; material to the right (left) of the breakpoint is inherited from the parent connected by the edge labelled $S$ ($P$) for ``suffix'' (``prefix'').} \label{ARG1}
\end{figure}

Consider a population evolving through the replication, mutation, and recombination of genetic material within individuals, emerging from a common origin and living through multiple generations until the present day. In general, the history of shared ancestry, mutation, and recombination events are not observed, and must be inferred from a sample of genetic data obtained from the present-day population. Crossover recombination can occur anywhere along a sequence, and the breakpoint position is also unobserved. This article focuses on methods for reconstructing possible histories of such a sample, in the form of \emph{ancestral recombination graphs (ARGs)} --- networks of evolution connecting the sampled individuals to shared ancestors in the past through coalescence, mutation, and crossover recombination events; an example is illustrated in Figure \ref{ARG1}. This is a very important but challenging problem, as many possible histories might have generated a given sample. Moreover, recombination can be undetectable unless mutations appear on specific branches of the genealogy \citep[Section 5.11]{heinbook}, and recombination events can produce patterns in the data that are indistinguishable from the effects of \emph{recurrent mutation} \citep{mcvean}; that is, two or more mutation events in a genealogical history that affect the same locus.

Parsimony is an approach focused on finding possible histories which minimise the number of recombinations and recurrent mutations. This does not necessarily describe the most biologically plausible version of events, but produces a useful lower bound on the complexity of the evolutionary pathway that might have generated the given dataset. Beyond specifying the types of events that are allowed, parsimony does not require assuming a particular generative model; the approach focuses on sequences of events that can generate the observed dataset, disregarding the timing and prior rate of these events. 

Previous work on reconstructing histories using parsimony has tackled recombination and recurrent mutation separately. Algorithms for reconstructing minimal ARGs generally make the \emph{infinite sites assumption}, which allows at most one mutation to have occurred at each site of the genome, thus precluding recurrent mutation events, and the goal is to calculate the minimum number of crossover recombinations required to explain a dataset, denoted $R_{min}$. Even with this constraint, the problem is NP-hard \citep{wang01}; exact algorithms are practical only for small datasets \citep{hein90,beagle}, and general methods rely on heuristic approximations \citep{hein93heuristic,shrub,margarita,paridaARG,thao}. Alternatively, one can assume the absence of recombination and seek to calculate the minimum number of recurrent mutations required, denoted $P_{min}$. In this case, reconstruction of maximum parsimony trees is also NP-hard \citep{foulds}; likewise, methods can only handle small datasets or are based on heuristics \citep[Section 5.4]{phylogenetics}.

Parsimony contrasts with the alternative approach of model-based inference, which requires the user to select a generative model and relies on the estimation of mutation and recombination rates as model parameters. Model-based inference generally involves integrating over the space of possible histories, which is usually intractable; methods rely on MCMC \citep[e.g.][]{argweaver} or importance sampling \citep[e.g.][]{jenkins_griffiths_11}, but the problem remains computationally difficult. If the presence of recombination is certain and reasonable models of population dynamics are available, model-based approaches may be more suitable and result in more powerful inference. However, model misspecification can play an important role, for instance when modelling viral evolution over a transmission network, where the relative importance of factors such as geographical structure, social clustering, and the impact of interventions may be difficult to ascertain. In this case, model-based inference can provide misleading results if overinterpreted, with poor quantification of uncertainty due to model misspecification. Parsimony-based methods fail to offer the interpretability or uncertainty quantification of a model but this does preclude their results being overinterpreted. They are simple and straightforward to implement and can be useful in situations such as enabling testing for the presence or absence of recombination when this is not certain \citep{bruen06}.

There are a number of recently developed methods, namely RENT+ \citep{rentp}, tsinfer \citep{tsinfer}, and Relate \citep{relate}, that seek to reconstruct local tree or ARG topologies from the data. These methods do not make strict model-based assumptions, incorporating heuristic algorithms, and do not aim to reconstruct the most \emph{parsimonious} histories. We note also the existence of numerous other methods for inference of recombination \citep[e.g.\ ][]{rdp, listephens, gard, 3seq} which do not explicitly reconstruct ARGs.

KwARG (``quick ARG") is a software tool, written in C, which implements a greedy heuristic-based parsimony algorithm for reconstructing histories that are minimal or near-minimal in the number of posited recombination and mutation events. The algorithm starts with the input dataset and generates plausible histories backwards in time, adding coalescence, mutation, recombination, and recurrent mutation events to reduce the dataset until the common ancestor is reached. By tuning a set of cost parameters for each event type, KwARG can find solutions consisting only of recombinations (giving an upper bound on $R_{min}$), only of recurrent mutations (giving an upper bound on $P_{min}$), or a combination of both event types. KwARG handles both the `infinite sites' and `maximum parsimony' scenarios, as well as interpolating between these two cases by allowing recombinations as well as recurrent mutations and sequencing errors, which is not offered by existing methods. This is illustrated in Figure \ref{ARG1}: KwARG finds all three types of solution for the given dataset. KwARG shows excellent performance when benchmarked against exact methods on small datasets, and outperforms existing parsimony-based heuristic methods on large, more complex datasets while maintaining computational efficiency; KwARG also achieves very good accuracy in reconstructing local tree topologies. The source code and executables are made freely available on GitHub at \url{https://github.com/a-ignatieva/kwarg}, along with documentation and usage examples. 

The paper is structured as follows. Details of the algorithm underlying KwARG are given in Section \ref{tech_details}, with an explanation of the required inputs and expected outputs. In Section \ref{performance}, the performance of KwARG on simulated data is benchmarked against exact methods and existing programs. An application of KwARG to a widely studied \emph{Drosophila melanogaster} dataset \citep{kreitman} is described in Section \ref{kreitman}. Discussion follows in Section \ref{discussion}. 

\section{Technical details} \label{tech_details}

Consider a sample of genetic data, where the allele at each site can be denoted 0 or 1. We do not make the infinite sites assumption, so that each site can undergo multiple mutation events. However, we do assume that mutations correspond to transitions between exactly two possible states, excluding for instance triallelic sites.

\subsection{Input}
KwARG accepts data in the form of a binary matrix, or a multiple alignment in nucleotide or amino acid format. The sequence and site labels can be provided if desired. It is possible to specify a root sequence, or leave this to be determined. The presence of missing data is permitted; regardless of the type of input, the data is converted to a binary matrix $\D$, with entries `$\star$' denoting missing entries or material that is not ancestral to the sample.

\subsection{Methods}
Under the infinite sites assumption, at most one mutation is allowed to have occurred per site. If any two columns contain all four of the configurations 00, 01, 10, 11, then the data could not have been generated only through replication and mutation, and there must have been at least one recombination event between the two corresponding sites. This is the four gamete test \citep{hudsonkaplan85}, and the two sites are said to be \emph{incompatible}. When recurrent mutations are allowed, the incompatibility could likewise have been generated through multiple mutations affecting the same site \citep{mcvean}.

KwARG reconstructs the history of a sample backwards in time, by starting with the data matrix $\D$ and performing row and column operations corresponding to coalescence, mutation, and recombination events, until only one ancestral sequence remains. By reversing the order of the steps, a forward-in-time history is obtained, showing how the population evolved from the ancestor to the present sample. When a choice can be made between multiple possible events, a neighbourhood of candidate ancestral states is constructed, using the same general method as that employed in the program Beagle \citep{beagle}. A backwards-in-time approach has also been implemented in the programs SHRUB \citep{shrub}, Margarita \citep{margarita} and GAMARG \citep{thao}, all of which adopt the infinite sites assumption but use different criteria for choosing amongst possible recombination events.

\subsubsection{Construction of a history} \label{clean}

For convenience, assume that the all-zero sequence is specified as the root, and 0 (1) entries of $\mathcal{D}$ correspond to ancestral (mutated) sites. Suppose $\D_t$ is the data matrix obtained after $t-1$ iterations of the algorithm. At the beginning of the $t$-th step, KwARG first reduces $\D_t$, by repeatedly applying the `Clean' algorithm \citep{wabi} through:
\begin{itemize} 
\item deleting uninformative columns (consisting of all 0's);
\item deleting columns containing only one 1 (corresponding to ``undoing" a mutation present in only one sequence);
\item deleting a row if it agrees with another row (corresponding to a coalescence event);
\item deleting a column if it agrees with an adjacent column.
\end{itemize}
Two rows (columns) \emph{agree} if they are equal at all positions where both rows (columns) contain ancestral material, and the sites (sequences) carrying ancestral material in one are a subset of the sites (sequences) carrying ancestral material in the other. 

A run of the `Clean' algorithm repeatedly applies these steps to $\D_t$, terminating when no further reduction is possible. Suppose the resulting data matrix is $\bD_t$. KwARG then constructs a neighbourhood $\N_t$ of candidate next states, each one obtained through one of the following operations:
\begin{itemize}
\item Pick a row and split it into two at a possible recombination point. Only a subset of possible recombining sequences and breakpoints needs to be considered; see \citet[Section 3.3]{beagle} for a detailed explanation.
\item Remove a recurrent mutation, by selecting a column and changing a 0 entry to 1, or a 1 entry to 0. This is the event type that is disallowed by algorithms applying the infinite sites assumption.
\end{itemize}
Suppose a neighbourhood $\N_t = \{\N_t^1, \hdots, \N_t^{N}\}$ is formed, consisting of all possible states that can be reached from $\bD_t$ through applying one of these operations. Then the reduced neighbourhood $\bN_t = \{ \bN_t^1, \hdots, \bN_t^N \}$ is formed by applying `Clean' to each state in turn. Each state $\bN_t^i$ is then assigned a score $S(\bN_t^i, \N_t^i, \bD_t)$, combining (i) the cost $C\left(\N_t^i, \bD_t \right)$, defined below, of reaching the configuration $\N_t^i$ from $\bD_t$, (ii) a measure $\am \left( \bN_t^i \right)$ of the complexity of the resulting data matrix $\bN_t^i$, and (iii) a lower bound $L(\bN_t^i)$ on the remaining number of recombination and recurrent mutation events still required to reach the ancestral sequence from $\bN_t^i$. Finally, a state is selected, say $\bN_t^j$, based on its score, and we set $\D_{t+1} = \bN_t^j$. The process of reducing the dataset followed by constructing a neighbourhood and choosing the best move is repeated, until all incompatibilities are resolved and the root sequence is reached. Pseudocode for the `Clean' algorithm and KwARG is given in Supplementary Section \ref{code}.

\begin{figure}[htbp!]
\centering
        \includegraphics[width=\linewidth]{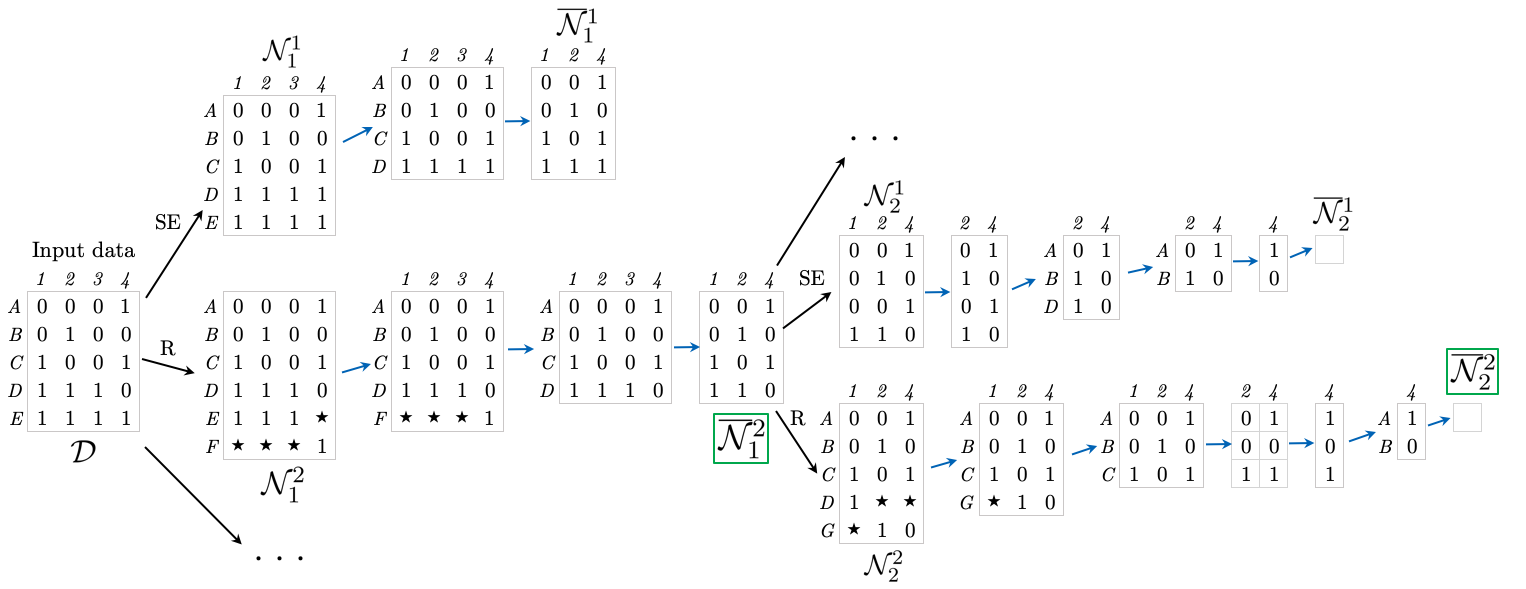}
    \caption{Example of a reconstructed history for the dataset in Figure \ref{ARG1}. Stars `$\star$' denote non-ancestral material. SE: recurrent mutation occurring on a terminal branch of the ARG. R: recombination event. A sequence of blue arrows corresponds to one application of the `Clean' algorithm. Green boxes highlight the selected states.} \label{arg_history}
\end{figure}

The construction of a history for the dataset given in Figure \ref{ARG1} is illustrated in Figure \ref{arg_history}. The first step corresponds to the construction of a neighbourhood, two of the states $\N_1^1, \N_1^2 \in \N_1$ are pictured. Then, the `Clean' algorithm is applied to each state in the neighbourhood (illustrated as a series of steps following blue arrows). From the resulting reduced neighbourhood $\{ \bN_1^1, \bN_1^2, \hdots \}$, the state $\bN_1^2$ is selected; the other illustrated path is abandoned. This process is repeated until all incompatibilities are resolved and the empty state is reached. Following the path of selected moves in this figure left-to-right corresponds to the events encountered when traversing the leftmost ARG in Figure \ref{ARG1} from the bottom up. If instead the state $\bN_2^1$ were selected at the second step of the algorithm, the resulting path would correspond to the ARG in the centre of Figure \ref{ARG1}. 

\subsubsection{Score}
When considering which next step to take, more informed choices can be made by considering not just the cost of the step, but also the complexity of the configuration it leads to. This is the principle behind the A* algorithm \citep{hart}, using a heuristic estimate of remaining distance to guide the choice of the next node to expand. KwARG applies the same principle in a greedy fashion, following a path of locally optimal choices in an attempt to find a minimal history.

The score implemented in KwARG is 
\begin{equation} \label{score_eq}
S\left(\bN_t^i, \N_t^i, \bD_t\right) = \left(C\left(\N_t^i, \bD_t \right)+ L\left(\bN_t^i\right)\right) \cdot \maxam\left(\bN_t\right) + \am\left(\bN_t^i\right),
\end{equation}
where
\[
L(\bN_t^i) = \begin{cases}
R_{min}\left(\bN_t^i \right)  &\text{if } \maxam(\bN_t) < 75,\\
HB\left(\bN_t^i \right) &\text{if } 75 \leq \maxam(\bN_t) < 200,\\
HK\left(\bN_t^i\right) &\text{otherwise.}
\end{cases}
\]
Here, $C\left(\N_t^i, \bD_t \right)$ denotes the cost of the corresponding event, defined in Section \ref{event_cost}; $\maxam(\bN_t)$ denotes the maximum amount of ancestral material seen in any of the states in $\bN_t$, and $\am(\bN_t^i)$ gives the amount of ancestral material in state $\bN_t^i$. Incorporating a measure of the amount of ancestral material in a state helps to break ties by assigning a smaller score to simpler configurations.

The method of computing the lower bound $L$ depends on the complexity of the dataset, with a trade-off between accuracy and computational cost. For relatively small datasets, it is feasible to compute $R_{min}$ exactly using Beagle. $HB$ refers to the haplotype bound, employing the improvements afforded by first calculating local bounds for incompatible intervals, and applying a composition method to obtain a global bound \citep{myers_griffiths}. $HK$ refers to the Hudson-Kaplan bound \citep{hudsonkaplan85}; this is quick but less accurate, so is reserved for larger,  more complex configurations. Note that these bounds are computed under the infinite sites assumption. 

The particular form and components of the score were chosen through simulation testing; we found that the given formula provides a good level of informativeness regarding the quality of a possible state.

\subsubsection{Event cost} \label{event_cost}
Each type of event is assigned a cost, which gives a relative measure of preference for each event type in the reconstructed history:
\begin{itemize}
\item $C_R$: the cost of a single recombination event, defaults to 1.
\item $C_{RR}$: the cost of performing two successive recombinations, defaults to 2. It is sufficient to consider at most two consecutive recombination events before a coalescence \citep{beagle}; this type of event also captures the effects of gene conversion.
\item $C_{RM}$: the cost  of a recurrent mutation. If $\N_t^i$ is formed from $\bD_t$ by a recurrent mutation in a column representing $k$ agreeing sites, this corresponds to proposing $k$ recurrent mutation events, so the cost is $C(\N_t^i, \bD_t) = k \cdot C_{RM}$. 
\item $C_{SE}$: this event is a recurrent mutation which affects only one sequence in the original dataset, i.e.\ it occurs on the terminal branches of the ARG. Thus, the event can be either a regular recurrent mutation, or an artefact due to sequencing errors. The cost can be set to equal $C_{RM}$, or lower if the presence of sequencing errors is considered likely.
\end{itemize}

KwARG allows the specification of a range of event costs as tuning parameters, as well as the number $Q$ of independent runs of the algorithm to perform for each cost configuration. The proportions of recombinations to recurrent mutations in the solutions produced by KwARG can be controlled by varying the ratio of costs for the corresponding event types. 

\subsubsection{Selection probability}
The method of selecting the next state from a neighbourhood of candidates will impact on the efficiency and performance of the algorithm. At one extreme, selecting at random amongst the states will mean that the solution space is explored more fully, but will be prohibitively inefficient in terms of the number of runs needed to find a near-optimal solution. On the other hand, always greedily selecting the move with the minimal score will quickly identify a small set of solutions for each cost configuration, at the expense of placing our faith in the ability of the score to assess the quality of the candidate states accurately.

We propose a selection method that is intermediate between these two extremes, randomising the selection but focusing on moves with near-minimal scores. A pseudo-score for state $\bN_t^i$ is calculated:
\begin{equation} \label{ann_eq}
\exp(T \cdot \left(1 - \widetilde{S}\left(\bN_t^i, \N_t^i, \bD_t\right)\right)),
\end{equation}
where
\begin{equation*}
\widetilde S\left(\bN_t^i, \N_t^i, \bD_t\right) = \frac{S\left(\bN_t^i, \N_t^i, \bD_t\right)- \min_j S\left(\bN_t^j, \N_t^j, \bD_t\right) }{\max_j S\left(\bN_t^j, \N_t^j, \bD_t\right) - \min_j S\left(\bN_t^j, \N_t^j, \bD_t\right)},
\end{equation*}
and states in $\mybar{\N_t}$ are selected with probability proportional to their pseudo-score. The annealing parameter $T$ controls the extent of random exploration; $T=0$ corresponds to choosing uniformly at random from the neighbourhood of candidates, and $T=\infty$ to always choosing a state with the minimal score. The default value of $T=30$ was chosen following simulation testing, which showed that this provides a good balance between efficiency and thorough exploration of the neighbourhood.

\subsection{Output}
The default output consists of the number of recombinations and recurrent mutations in each identified solution; an example for the Kreitman dataset is given in Table \ref{table2}. Each iteration is assigned a unique random seed, which can be used to reconstruct each particular solution and produce more detailed outputs, such as a detailed list of events in the history, the ARG in several graph formats, or the corresponding sequence of marginal trees. 

\section{Performance on simulated data} \label{performance}

We have tested the performance of KwARG based on two main criteria. Firstly, we compared its performance against exact methods, PAUP* and Beagle, to demonstrate that KwARG successfully reconstructs minimal histories in the mutation-only and recombination-only cases, respectively. Secondly, we carried out simulation studies to determine how accurately KwARG reconstructs local trees, compared against three other methods: tsinfer, RENT+, and ARGweaver \citep{argweaver}. Finally, we compared how well KwARG performs against the parsimony-based heuristic methods SHRUB \citep{shrub} and SHRUB-GC \citep{shrub_gc}; these results are presented in Supplementary Section \ref{app2}. We also investigated the dependence of the run time of KwARG on the number and length of sequences, through simulation studies.

\subsection{Finite sites}

\subsubsection{Comparison to PAUP*}
Disallowing recombination, the quality of computed upper bounds on $P_{min}$ was tested by comparison with PAUP* \citep[version 4.0a168]{paup}, which was used to compute the exact minimum parsimony score via branch-and-bound on 994 datasets simulated as described in Supplementary Section \ref{paup_details}.

KwARG failed to find $P_{min}$ in 11 (1.1\%) cases out of 994. The results are illustrated in the left panel of Figure \ref{recomb_results}. Where KwARG failed to find an optimal solution, in all 11 cases it was off by just one recurrent mutation. Figure \ref{recomb_results} also demonstrates that a substantial proportion of recurrent mutations do not create incompatibilities in the data, and the number of actual events often far exceeds $P_{min}$.

\subsection{Infinite sites}

\subsubsection{Comparison to Beagle}
Under the infinite sites assumption (disallowing recurrent mutation), the accuracy of KwARG's upper bound on $R_{min}$ was tested by comparison with Beagle \citep{beagle}, on 1\,037 datasets simulated as described in Supplementary Section \ref{beagle_details}. 

Using the default annealing parameter $T=30$, KwARG found $R_{min}$ in all cases. In 97\% of the runs, this took under 5 seconds of CPU time (on a 2.7GHz Intel Core i7 processor); all but one run took less than 40 seconds. In 93\% of the runs, 1 iteration was sufficient to find an optimal solution; in 99\% of the runs, 5 iterations were sufficient. Beagle found the exact solution in 5 seconds or less in 86\% of cases; for datasets with a small $R_{min}$ Beagle runs relatively quickly (median run time for $R_{min} = 5$ was 1 second, compared to KwARG's 0.3 seconds). For more complex datasets, KwARG finds an optimal solution much faster; for $R_{min} = 9$, the median run time of Beagle was 56 seconds, compared to KwARG's 3 seconds.

Setting $T=10$ and $T=\infty$ resulted in 5 and 22 failures to find an optimal solution, respectively, when KwARG was run for $Q=1\,000$ iterations per dataset (or terminated after 10 minutes have elapsed), demonstrating that setting the annealing parameters too low or too high results in deterioration of performance.

\begin{figure}[htbp!]
\centering
        \includegraphics[height=7.6cm]{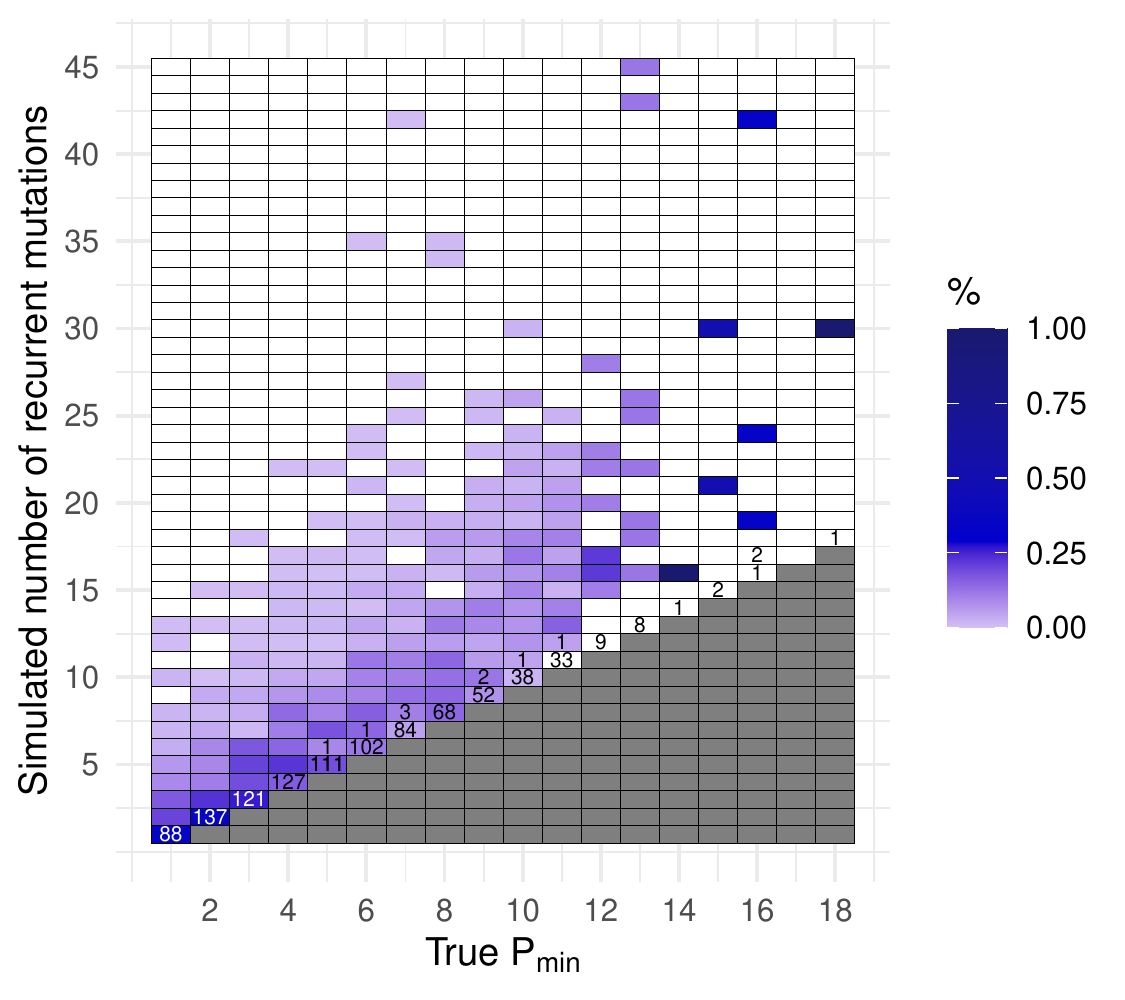}
        \includegraphics[height=7.6cm]{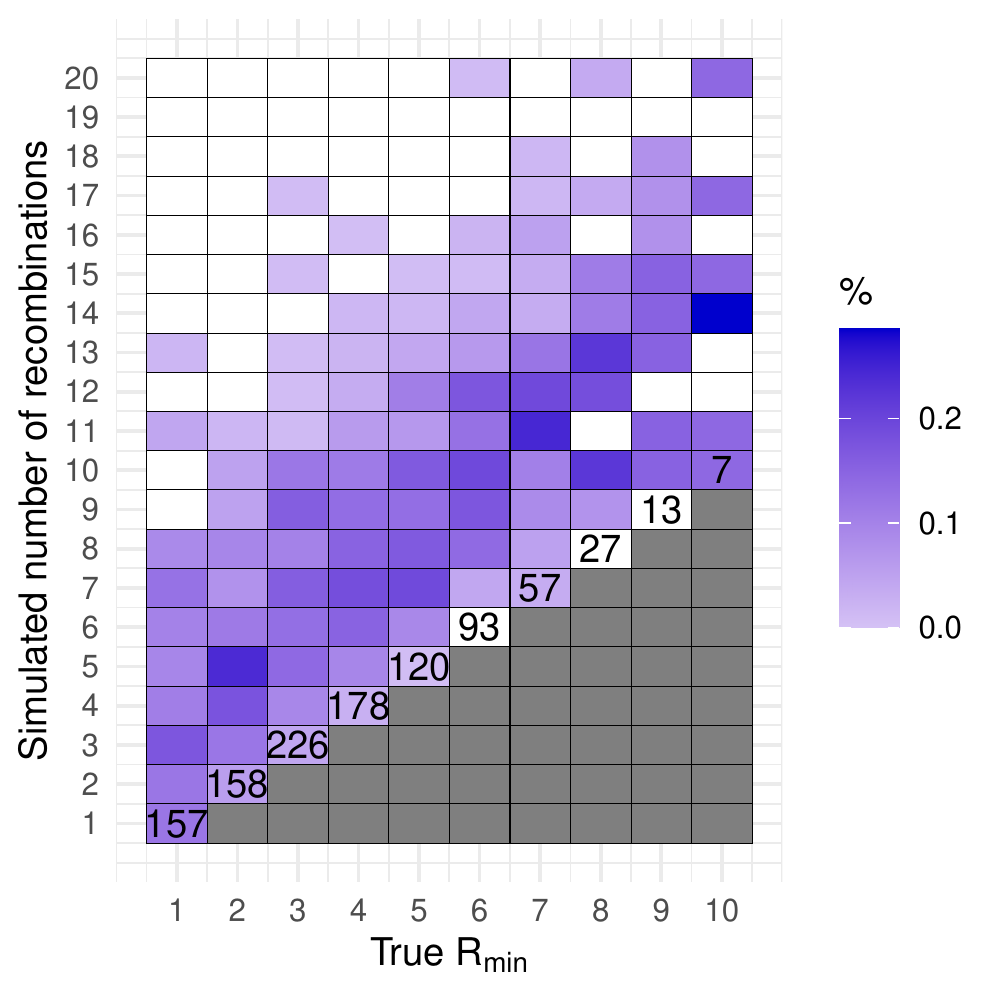}
        \setlength{\abovecaptionskip}{4pt}
    \caption{Left: number of simulated recurrent mutations against $P_{min}$. Right: number of simulated recombinations against $R_{min}$. Cell colouring intensity is proportional to the number of datasets generated for each pair of coordinates. Numbers in each cell correspond to the number of cases where for a dataset with the true minimum number of events given on the $x$-axis, KwARG inferred the number of events given on the $y$-axis (unlabelled cells correspond to 0 such cases).} \label{recomb_results}
\end{figure}

The right panel of Figure \ref{recomb_results} illustrates the results, and shows the relationship between the true simulated number of recombinations and $R_{min}$. This demonstrates that in many cases, substantially more recombinations have occurred than can be confidently detected from the data. 

\subsubsection{Comparison to tsinfer, RENT+, and ARGweaver.}

We tested the performance of KwARG in recovering the topology of simulated local trees for a range of recombination and mutation rates (under the infinite sites assumption). For each combination of rates, we simulated 100 datasets; details of the simulation parameters and settings used in running each program are given in Supplementary Section \ref{app_comparisons_s}. From the output of each method, we calculated the Kendall--Colijn metric \citep{kc_metric} between the inferred and true tree topologies at each variant site position, calculating the mean across all variant sites and averaging over the 100 datasets. We note that ARGs contain more information than local trees, but there is no obvious way of comparing ARG topologies (and tsinfer only infers local trees, rather than full ARGs). 

\begin{figure}[htbp!]
\centering
        \includegraphics[width=0.4\linewidth]{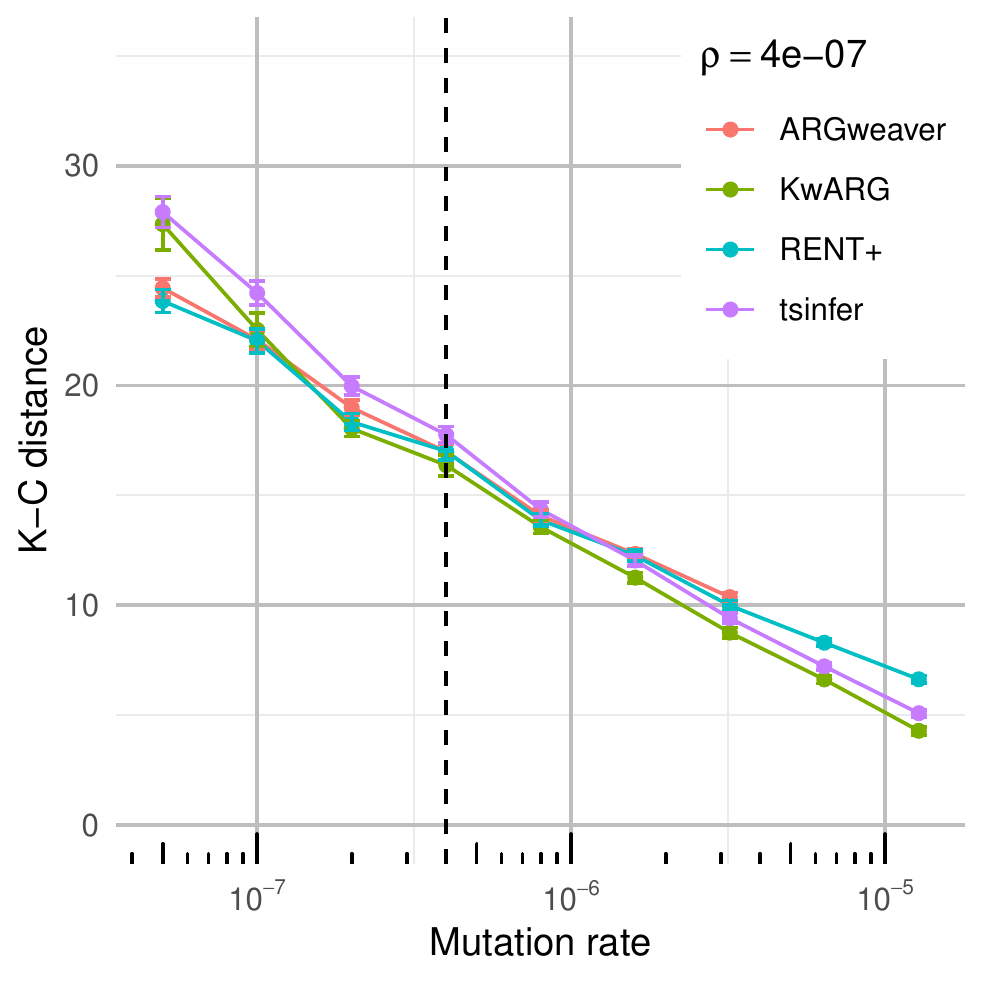}
        \includegraphics[width=0.4\linewidth]{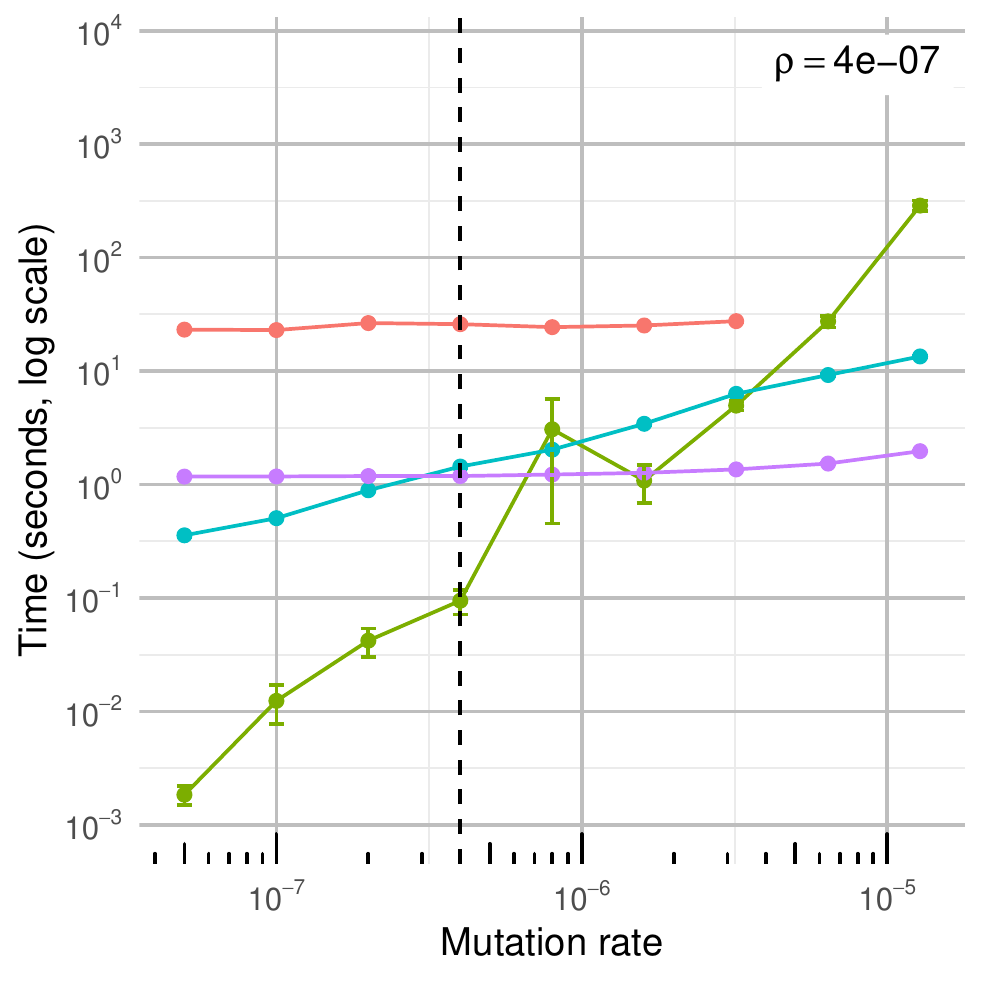}
    \caption{Comparison of performance in inferring local trees. Left panel: points show mean across 100 simulated datasets for each value of mutation rate $\mu$ (per generation per site) with recombination rate $\rho = 4 \cdot 10^{-7}$ (per generation per site); error bars show mean $\pm$ standard error. Lower K-C distance indicates better accuracy. Right panel: points show mean run time averaged over 100 datasets for each combination of rate parameters; error bars show mean $\pm$ standard error. ARGweaver results not shown past $\mu = 3.2 \cdot 10^{-6}$ due to prohibitively long run time.} \label{comparisons}
\end{figure}

The results are shown in the left panel of Figure \ref{comparisons} and Supplementary Figure \ref{app_comparisons}. All methods show very comparable performance across the range of considered scenarios, with KwARG slightly outperforming the other methods, based on the chosen metric, when the recombination rate is relatively low and the mutation rate relatively high. We have performed the same analysis using the Robinson--Foulds metric \citep{RF_metric}, and found this to give very similar results.

\subsection{Run time analysis}

A comparison of the run times of KwARG against tsinfer, RENT+, and ARGweaver is presented in the right panel of Figure \ref{comparisons} and Supplementary Figure \ref{app_comparisons_time}. KwARG demonstrates good efficiency when the recombination and mutation rates are relatively low, and shows roughly linear growth in run time as the mutation rate increases.

The dependence of the run time of KwARG on the number and length of sequences was further investigated through simulations; the results are presented in Supplementary Section \ref{time}. Keeping the sequence length fixed showed that KwARG runs very quickly when the number of sequences is very low, and shows roughly exponential growth in run time when the number of sequences is $6$ or more. Keeping the number of sequences fixed shows that, after an initial exponential increase (due to small datasets taking very little time per iteration), the run time scales roughly linearly in sequence length.

\section{Application to Kreitman data} \label{kreitman}
\begin{figure}[htbp!]
\centering
        \includegraphics[width=0.92\linewidth]{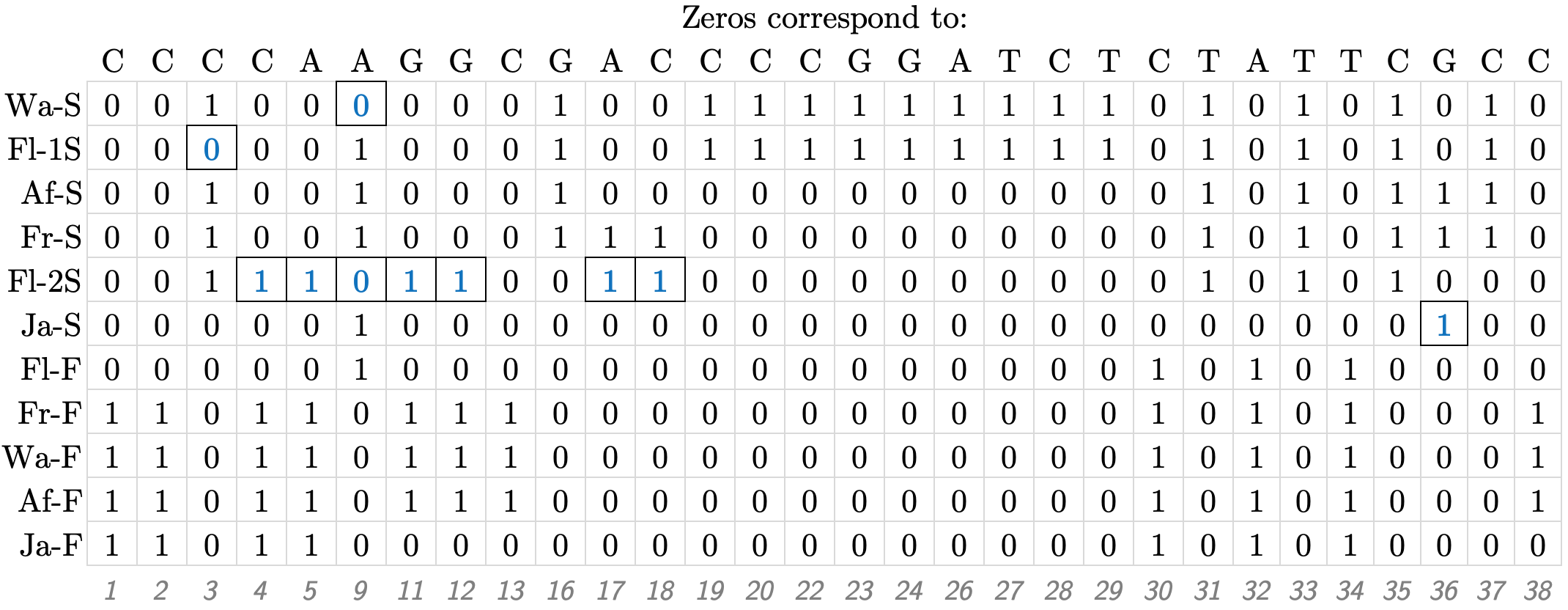}
    \caption{Illustration of the Kreitman dataset. The 11 sequences labelled as in \citet{kreitman}; polymorphic sites are labelled 1--43 and columns with singleton mutations are not shown.} \label{kreitman3}
\end{figure}

The performance of KwARG is illustrated on the classic dataset of \citet[Table 1]{kreitman}; this is not close to the performance limit of KwARG, but has been widely used for benchmarking algorithms used for ARG reconstruction. The dataset consists of 11 sequences and 2\,721 sites, of which 43 are polymorphic, of the alcohol dehydrogenase locus of \emph{Drosophila melanogaster}.  The data is shown in Figure \ref{kreitman3}, with columns containing singleton mutations removed for ease of viewing. Applying the `Clean' algorithm, as described in Section \ref{clean}, reduces this to matrix of 9 rows and 16 columns. KwARG was run with the default parameters, $Q=500$ times for each of 13 default cost configurations given in Supplementary Section \ref{app1}. An example of the output is shown in Table \ref{table2}. 

KwARG correctly identified the $R_{min}$ of 7 and the $P_{min}$ of 10 (confirmed by running Beagle and PAUP*, respectively). The 6\,500 iterations of KwARG took just under 9 minutes to run. Of these, 1,829 (28\%) resulted in optimal solutions; some are shown in Table \ref{table2}. KwARG identified multiple combinations of recombinations and recurrent mutations that could have generated this dataset. By default, slightly cheaper costs are assigned to recurrent mutations if they happen on terminal branches, so the results show a bias towards solutions with more $SE$ events for each given number of recombinations. 
\begin{table}[htbp!]
\begin{tabular}{cccrrrrcccc}
& Seed       & $T$ & $C_{SE}$ & $C_{RM}$ & $C_R$ & $C_{RR}$ & SE & RM & R & $\sum_{t} |\N_t|$\\ \cline{2-11}
1 & 2263536315  & 30.0  & $\infty$ &  $\infty$   &  1.00   &  2.00 &  0  & 0  & 7   &    143                             \\
 2 & 2347021759 &  30.0    & 0.90  &   0.91   &  1.00  &   2.00 &  1  & 0  & 6     &   853        \\
3 & 1791455164  & 30.0 &    0.80  &   0.81  &   1.00  &   2.00  & 1   &0  & 5    &    728      \\
4 & 1684879495  & 30.0 &    0.60   &  0.61  &   1.00  &   2.00  & 2  & 0 &  4    &    783      \\
5 & 1884182000 &  30.0  &   0.40   &  0.41  &   1.00   &  2.00  & 3 &  0 &  3    &    806                                \\
6 & 1900122424 &  30.0 &    0.20   &  0.21  &   1.00   &  2.00 &  5 &  0  & 2    &    702      \\
7 & 2111915557  & 30.0   &  0.10  &  0.11  &   1.00  &   2.00 &  8  & 0 &  1    &    833       \\
8 & 2888657821 &  30.0  &   0.01  &   0.02  &   1.00   &  2.00 & 10 &  0 &  0   &     715      \\                 
\end{tabular}
\vspace{5pt}
\caption{Example output of KwARG for the Kreitman dataset. SE: number of recurrent mutations occurring on terminal branches of the ARG (possible sequencing errors). RM: number of other recurrent mutations. R: number of recombinations. Last column gives the total number of neighbourhood states considered.} \label{table2}
\end{table}

\begin{figure}[htbp!]
\centering
        \includegraphics[width=0.85\linewidth]{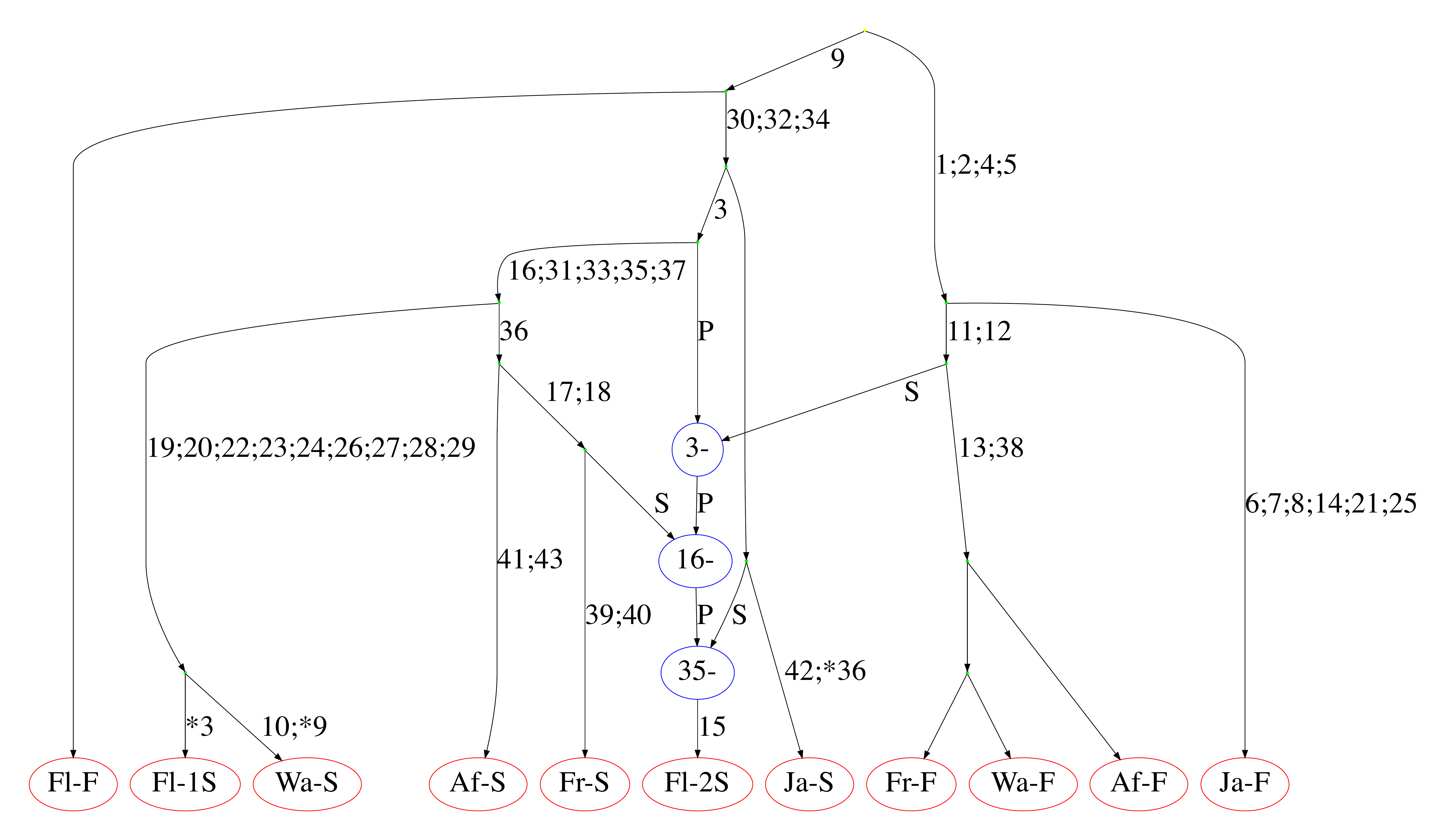}
    \caption{ARG constructed for the Kreitman data. Edges are labelled with sites undergoing mutations; recurrent mutations are prefixed with an asterisk. Recombination nodes, in blue, are labelled with the recombination breakpoint; material to the right (left) of the breakpoint is inherited from the parent connected by the edge labelled $S$ ($P$) for ``suffix'' (``prefix''). } \label{ARG2}
\end{figure}

The ten recurrent mutations appearing in the solution in row 8 of Table \ref{table2} are highlighted on the dataset in Figure \ref{kreitman3}. It is striking that 7 of these 10 recurrent mutations affect the same sequence Fl-2S. In fact, these 7 recurrent mutations could be replaced by 3 recombination events affecting sequence Fl-2S, with breakpoints just after sites 3, 16, and 35; leaving the other identified recurrent mutations unchanged yields the solution in row 5 of Table \ref{table2}. These findings suggest that the sequence may have been affected by cross-contamination or other errors during the sequencing process, or it could indeed be a recombinant mosaic of four other sequences in the sample. This recovers the results obtained by \citet{stephensdrosophila}, who posited the recombinant origins of sequence Fl-2S following manual examination of a reconstructed maximum parsimony tree, which also highlighted the five consecutive mutations identified by KwARG. The ARG corresponding to the solution in row 5 of Table \ref{table2}, visualised using Graphviz \citep{graphviz}, is shown in Figure \ref{ARG2}.

Examination of the identified solutions also shows that site 36 of sequence Ja-S ``necessitates" two of the seven recombinations inferred in the minimal solution in the absence of recurrent mutation, while sites 3 and 9 in sequences Wa-S and Fl-1S, respectively, each create incompatibilities that could be resolved by one recombination.

\section{Discussion} \label{discussion}

Methods for the reconstruction of parsimonious ARGs generally rely on the infinite sites assumption. When examining the output ARGs, it is often difficult to tell by how much the inferred recombination events actually affect the recombining sequences. As is the case with the Kreitman dataset, sometimes further examination reveals that two crossover recombination events have the same effect as one recurrent mutation, raising questions about which version of events is more likely. KwARG removes the need for such manual examination, and provides an automated way of highlighting such cases, which is particularly useful for larger datasets. 

While KwARG performs well in inferring ARGs under the infinite sites assumption, it can be particularly useful in analysing genetic data from organisms whose genomes are reasonably likely to undergo recurrent mutation, such as viruses with relatively high mutation rates and short genomes. One such application is demonstrated in \citet{sarscov2_paper}, where the output of KwARG is combined with probabilistic arguments to investigate the presence of ongoing recombination in SARS-CoV-2. 

The solutions identified by KwARG differ in the proportion of recurrent mutations to recombinations, ranging from an explanation that invokes only recombination events to one that invokes only mutation events. As is the case with other heuristic and parsimony-based methods, KwARG cannot offer uncertainty quantification for the inferred ARGs. Quantifying the likelihood of each scenario will be application-specific; for instance, one can choose a reasonable model of evolution for the population being studied, and identify the most likely solution under a range of reasonable mutation and recombination rates. When the presence or absence of recombination is not certain, then should the number of recurrent mutations needed to explain the dataset be infeasibly large, this provides evidence for the presence of recombination; this is the idea underlying the homoplasy test of \citet{homoplasytest}. If the largest ``reasonable" number of recurrent mutations is then estimated, KwARG can be used to say how many additional recombination events are required to explain the dataset.

KwARG performs well when compared against exact parsimony methods for the `recombination-only' and `mutation-only' scenarios. Because of the random exploration incorporated within KwARG, it should be run multiple times on the same dataset before selecting the best solutions; the optimal run length of KwARG will be constrained by timing and the available computational resources. To gauge whether KwARG has run enough iterations, one could proceed by calculating $R_{min}$ and $P_{min}$ either exactly (if the data is reasonably small) or using other heuristics-based methods (such as SHRUB or PAUP*), to confirm whether KwARG has found good solutions at these two extremes. 

The range of solutions explored by KwARG is guided by the choice of cost parameters. As a rule of thumb, simulations have shown that if the mutation and recombination rates are similar, costs near one give good accuracy of solutions in terms of reconstructing local tree topologies; if the mutation rate is significantly higher (lower) than the recombination rate, the cost should be set to less than (greater than) one. As KwARG incorporates a degree of random exploration, a range of solutions will still be obtained; the best choice of parameters will depend strongly on the nature and aims of the analysis being performed. 

For model-based inference, the modelling assumptions can obviously affect the quality of the results; however, a parsimony-based approach also makes the strong assumption that the minimal ARG can capture useful information about the history of a sample. This will obviously depend strongly on the true recombination rate. Based on our comparisons with RENT+, tsinfer, and ARGweaver, KwARG achieves very good accuracy of inference of local tree topologies at least comparable to these other methods, particularly when the recombination rate is low to moderate and the mutation rate moderate to high. We emphasise that KwARG demonstrates relatively good accuracy even when the recombination rate is high and even though its express goal is to seek the most parsimonious, rather than necessarily the most likely, history. Moreover, for datasets with relatively few incompatibilities, the run time of KwARG is competitive with that of the other methods. It is also interesting to note that although all four programs incorporate very different approaches and heuristic algorithms, they demonstrate very similar performance in inferring local tree topologies over the range of considered scenarios.

The scalability of KwARG remains a challenge for large and more complex datasets. Performance gains could be readily achieved by running multiple iterations of KwARG in parallel, or incorporating more efficient ways of storing the intermediate states. Further improvements could also be obtained by amending the calculation of lower bounds within the cost function in order to account for the presence of recurrent mutation, which should make the scores more accurate, and hence the neighbourhood exploration more efficient. Other avenues for further work include explicitly incorporating gene conversion as a possible type of recombination event with a separate cost parameter, with a view to developing the underlying model of evolution to even more closely reflect biological reality.

\section{Acknowledgements}
We thank two anonymous reviewers for their helpful comments. This work was supported by the Engineering and Physical Sciences Research Council and the Medical Research Council through the OxWaSP Centre for Doctoral Training [EPSRC grant number EP/L016710/1], and by the Alan Turing Institute [EPSRC grant number EP/N510129/1].

\bibliography{literature}{}
\bibliographystyle{statsy}

\pagebreak
\nolinenumbers
\begin{center}
\textbf{\large Supplementary Materials}
\end{center}
\setcounter{equation}{0}
\setcounter{figure}{0}
\setcounter{table}{0}
\setcounter{section}{0}
\setcounter{page}{1}
\pagenumbering{roman}
\makeatletter
\renewcommand{\theequation}{S\arabic{equation}}
\renewcommand{\thefigure}{S\arabic{figure}}
\renewcommand{\thesection}{S\arabic{section}}
\renewcommand{\bibnumfmt}[1]{[S#1]}
\renewcommand{\citenumfont}[1]{S#1}

\singlespacing

\section{KwARG pseudocode} \label{code}
Let $\D$ be an input data matrix with entries 0, 1 or $\star$. Denote by $\D_{i,j}$ the entry of $\D$ at position $(i,j)$. Let $R_r(\D, i)$ and $R_c(\D, j)$ denote the resulting matrix when the $i$-th row or the $j$-th column of $\D$ is deleted, respectively. Let the history $\HH$ be a set storing all of the intermediate states visited on the path from $\D$ to the root of the ARG.

\begin{algorithm}[htbp!]
\SetAlgoLined
\KwIn{Dataset $\D$,  history $\HH$}
\KwOut{Reduced dataset $\bD$, updated history $\HH'$}
Initialise $C \gets$ true, $\bD \gets \D, \; \HH' \gets \HH$\;
\While{$C$}{
    \uIf{two distinct rows $i, j$ agree: $\bD_{i,k} \in \{ \bD_{j,k} \, , \star \} \; \forall k$}{
      $\bD \gets R_r(\bD, i), \; \HH' \gets \HH' \cup \bD$ \;
    }
    \uElseIf{there is a column $i$ such that $\bD_{k, i} = 1$ for exactly one $k$}{
      $\bD \gets R_c(\bD, i), \; \HH' \gets \HH' \cup \bD$ \;
    }
       \uElseIf{two distinct neighbouring columns $i, j$ agree:  $\bD_{k,i} \in \{ \bD_{k,j}\, , \star \} \; \forall k$}{
      $\bD \gets R_c(\bD, i), \; \HH' \gets \HH' \cup \bD$ \;
    }
        \uElse{
      $C \gets$ false\;
    }
}
\Return{$(\bD, \, \HH')$}\;
 \caption{Clean \citep[adapted from][]{wabi}}
\end{algorithm}

Define the following operations:
\begin{enumerate}
\item Recurrent mutation: $\widetilde{\D} = \text{RM}(\D, i, j)$ is the result of a recurrent mutation in row $i$ at column $j$; $\widetilde{\D}$ is obtained from $\D$ by changing the $(i,j)$-th entry from 0 to 1 or from 1 to 0.
\item Recombination: $\widetilde{\D} = \text{Rec}(\D, i, j)$ is the result of a recombination in row $i$ with breakpoint just after column $j$. Namely, $\widetilde{\D}$ is obtained from $\D$ by inserting a copy of the $i$-th row just below itself, and setting $\widetilde{\D}_{i,k} = \star \; \forall k \leq j $ and $\widetilde{\D}_{i+1,k} = \star \; \forall k > j$.
\item Two consecutive recombinations: $\widetilde{\D} = \text{RRec}(\D, i, j, k, l)$ is the result of performing two recombinations, in rows $i$ and $k$ with breakpoints at $j$ and $l$, respectively. 
\end{enumerate}

Note that for recombination events, not all row and column positions should to be considered, as some moves are guaranteed not to resolve any incompatibilities in the dataset. We apply the ideas detailed in \citet[Section 3.3]{beagle} to restrict the rows and breakpoints considered for recombination events. Suppose that as a result, $\mathcal{R}$ is the list of row and column indices $(i,j)$ to consider for recombination events, and $\mathcal{R}\mathcal{R}$ is the list of indices $(i,j,k,l)$ to consider for two consecutive recombination events.

\clearpage

\begin{algorithm}[htbp!]
\SetAlgoLined
\KwIn{Dataset $\D$}
\KwOut{Neighbourhood $\N$}
Initialise $\N \gets \{ \emptyset \}$\;
\For{$(i,j) \in \mathcal{R}$}{
		$\N \gets \N \cup \text{Rec}(\D, i, j)$\;
}
\For{$(i,j, k, l) \in \mathcal{R}\mathcal{R}$}{
		$\N \gets \N \cup \text{RRec}(\D, i, j, k, l)$\;
}
\For{all rows $i$}{
	\For{all columns $j$ such that $\D_{i,j} \neq \star$} {
		$\N \gets \N \cup  \text{RM}(\D, i, j)$\;
	}
}
\Return{$\N$}\;
\caption{Neighbourhood}
\end{algorithm}

\begin{algorithm}[htbp!]
\SetAlgoLined
\KwIn{Dataset $\D$}
\KwOut{History $\HH$}
Initialise $i \gets 1, \; \HH \gets \{ \D \}, \; (\bD_1, \HH) \gets Clean(\D, \HH)$\;
\While{$\bD_i  \neq \emptyset $}{
	$\bN_i \gets \{ \emptyset \}, \; \mathcal{L}_i \gets \{ \emptyset \}, \; S \gets \{ \emptyset \}$\;
	$\N_i \gets  Neighbourhood(\bD_i) = \{ \N_i^1, \N_i^2, \hdots \}$\;
	\For{$j = 1$ to $|\N_i|$} {
	$(\bN_i^j, \mathcal{L}_i^j) \gets Clean(\N_i^j, \HH \cup \N_i^j)$\;
	$\bN_i \gets \bN_i \cup \bN_i^j, \; \mathcal{L}_i \gets \mathcal{L}_i \cup \mathcal{L}_i^j $\;
	$S \gets S \cup \widetilde{S}\left(\bN_i^j, \N_i^j, \bD_i\right)$, where $\widetilde{S}\left(\bN_i^j, \N_i^j, \bD_i\right)$ is computed using \eqref{ann_eq}\;
	}
	Randomly draw an index $k$ from $\{1, \hdots, |\bN_i| \}$ with probabilities proportional to entries of $S$\;
	Set $\bD_{i+1} \gets \bN_i^k$, $\HH \gets \mathcal{L}_i^k$\;
	$i \gets i + 1$\; 
}
\Return{$\HH$}\;
 \caption{KwARG}
\end{algorithm}

\section{Default cost configuration} \label{app1}
If the number of iterations $Q>1$ is specified but no costs are input, KwARG runs each of the following 13 cost configurations $Q$ times: 
\begin{align*}
(C_{SE}, C_{RM}, C_R, C_{RR}) \in \{ &(\infty, \infty, 1.0, 2.0), (1.0, 1.01, 1.0, 2.0), (0.9, 0.91, 1.0, 2.0), (0.8, 0.81, 1.0, 2.0),\\
 &(0.7, 0.71, 1.0, 2.0), (0.6, 0.61, 1.0, 2.0), (0.5, 0.51, 1.0, 2.0), \\
 &(0.4, 0.41, 1.0, 2.0), (0.3, 0.31, 1.0, 2.0), (0.2, 0.21, 1.0, 2.0), \\
&(0.1, 0.11, 1.0, 2.0), (0.01, 0.02, 1.0, 2.0), (1.0, 1.1, \infty, \infty) \}.
\end{align*}

\begin{figure}[htbp!]
\centering
        \includegraphics[width=0.4\linewidth]{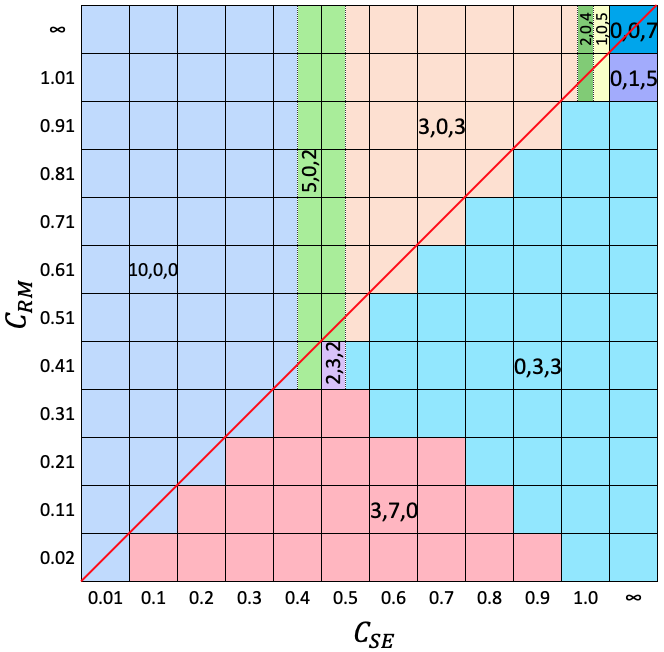}
    \caption{Solution tile plot for the Kreitman dataset.} \label{costs}
\end{figure}

The effectiveness of this is illustrated in Figure \ref{costs}, which is based on the set of all possible minimal solutions identified for the Kreitman dataset. Fixing $C_R = 1.0$ and $C_{RR} = 2.0$, each tile represents a pair $(C_{SE}, C_{RM})$. Each tile is coloured and labelled according to the corresponding cost-optimal solution, in the form $\{x, y, z\}$, giving the number of $SE$, $RM$ and recombination events, respectively. For instance, if $C_{SE} = 0.5$ and $C_{RM} = 0.61$, the solutions $\{3,0,3\}$ (with cost $3 \cdot 0.5 + 3 \cdot 1.0 = 4.5$) and $\{5,0,2\}$ (with cost $5 \cdot 0.5 + 2 \cdot 1.0 = 4.5$) have the lowest costs over all feasible solutions.

The default cost configuration includes all pairs $(C_{SE}, C_{RM})$ on the diagonal in this plot, falling on the red line. This line crosses all optimal solutions which maximise the number of $SE$ events for each possible number of recombinations. Such events affect only a single sequence at a single site in the input dataset, so are, in a sense, more parsimonious than recurrent mutations occurring on internal branches.

\section{Comparison to PAUP* and Beagle}

\subsection{PAUP*} \label{paup_details}
1\,100 genealogies were simulated using msprime \citep{msprime} (parameters: 20 sequences, $N_e = 1$). For each tree, Seq-Gen \citep{seqgen} was used to add mutations (parameters: 1\,000 sites, mutation rate per generation per site set by the scaling constant $s=0.01$); only transitions were allowed, to fulfil the requirement that sites mutate between exactly two states. 1\,063 datasets exhibited incompatibilities caused by recurrent mutations. KwARG was run for a total of $Q=600$ iterations per dataset; 150 of these were used to estimate $R_{min}$, and 450 were run with a range of costs to estimate $P_{min}$. The runs were terminated after 10 minutes (if 600 iterations had not been completed by then, the results were discarded; this happened in 69 cases); a total of 994 successful runs were performed.

\subsection{Beagle} \label{beagle_details}

1\,100 datasets were simulated using msprime, under the infinite sites assumption (parameters: $N_e = 1$, mutation rate per generation per site 0.02, recombination rate per site 0.0003, 40 sequences of length 2\,000bp). Of the generated datasets, 38 had no incompatible sites, and runs were terminated if Beagle took over 10 minutes to complete (which happened in 25 cases), leaving 1\,037 datasets for testing. The parameters were chosen to produce datasets on which Beagle could be run within a reasonable amount of time; the value of $R_{min}$ for the simulated datasets varied between 1 and 10. 

\section{Comparison to SHRUB and SHRUB-GC} \label{app2}

The performance of KwARG on larger datasets was tested against the parsimony-based heuristic methods SHRUB and SHRUB-GC. Both methods implement a backwards-in-time construction of ARGs, using a dynamic programming approach to choose among possible recombination events. SHRUB produces an upper bound on $R_{min}$ under the infinite sites assumption. SHRUB-GC also allows gene conversion events; setting the maximum gene conversion tract length to 1 makes this equivalent to recurrent mutation. The algorithm seeks to minimise the total number of events, essentially assigning equal costs to recombination and recurrent mutation. This differs from KwARG in that a single solution is produced for a given dataset, rather than a full range of solutions varying in the number of recombinations and recurrent mutations.

Using msprime and Seq-Gen, 300 datasets of 100 sequences were simulated, with a range of mutation and recombination rates and sequence lengths of 2\,000, 5\,000, 8\,000 and 10\,000 bp. For each dataset, KwARG was run for a total of $Q=260$ iterations, with the default cost configurations and  $T=30$. The resulting upper bound on $R_{min}$ was compared to that produced by SHRUB, and the minimum number of events over all identified solutions was compared to the solution produced by SHRUB-GC (configured to allow length-1 gene conversions).

\begin{figure}[htbp!]
\centering
        \includegraphics[width=0.35\linewidth]{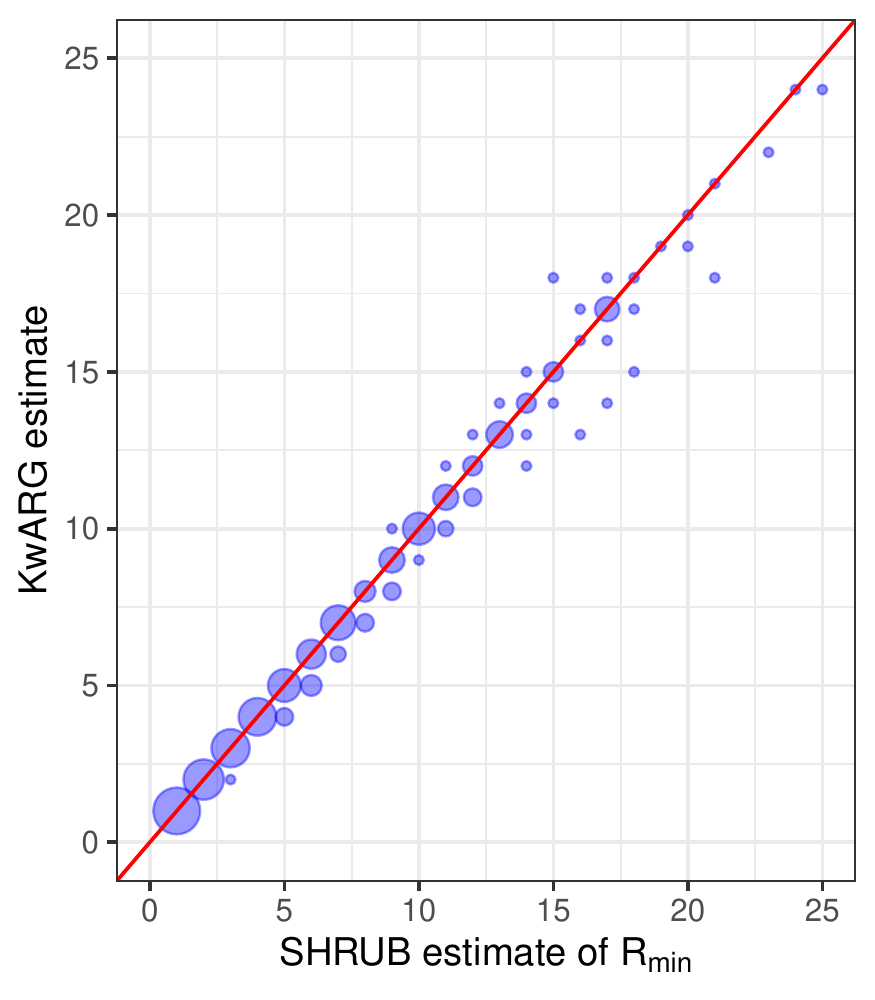}
        \includegraphics[width=0.35\linewidth]{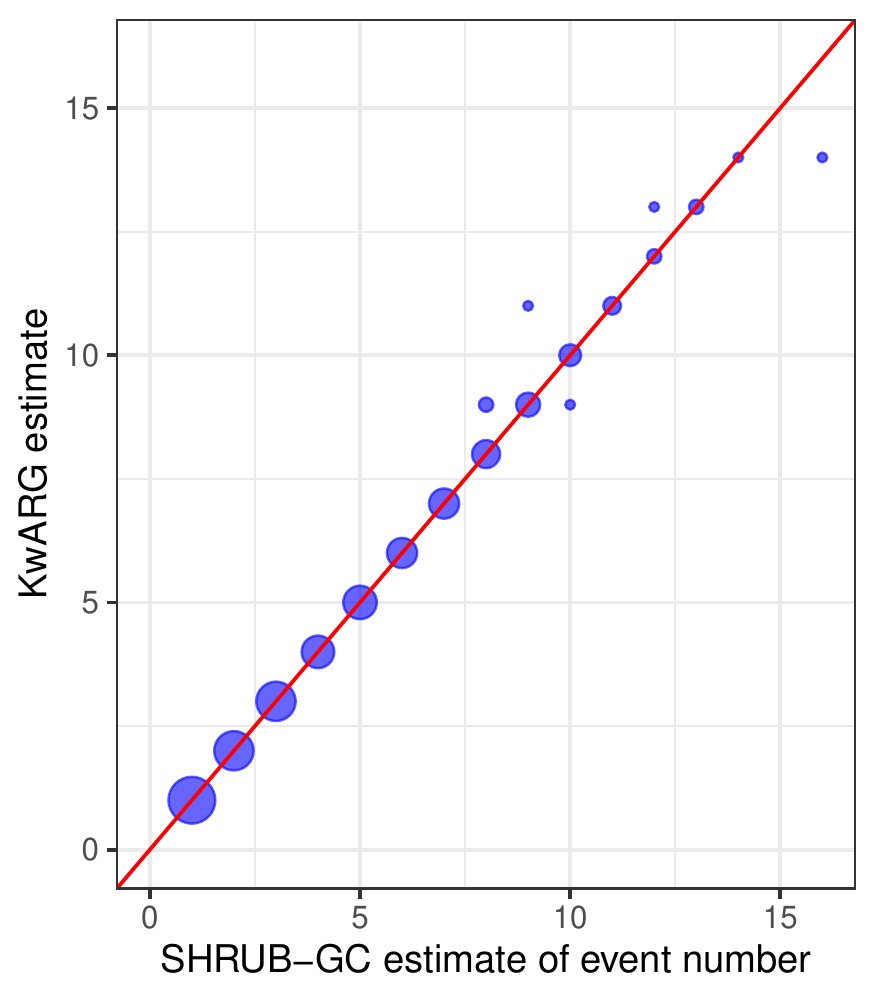}
    \caption{Comparison of KwARG to SHRUB and SHRUB-GC. $x$-axis: estimate produced by SHRUB (left) and SHRUB-GC (right). $y$-axis: estimate produced by KwARG. Instances where equally good solutions were found lie on the red diagonal line. Size of points is proportional to the number of corresponding datasets.} \label{shrub}
\end{figure}

\begin{figure}[htbp!]
\centering
        \includegraphics[width=0.4\linewidth]{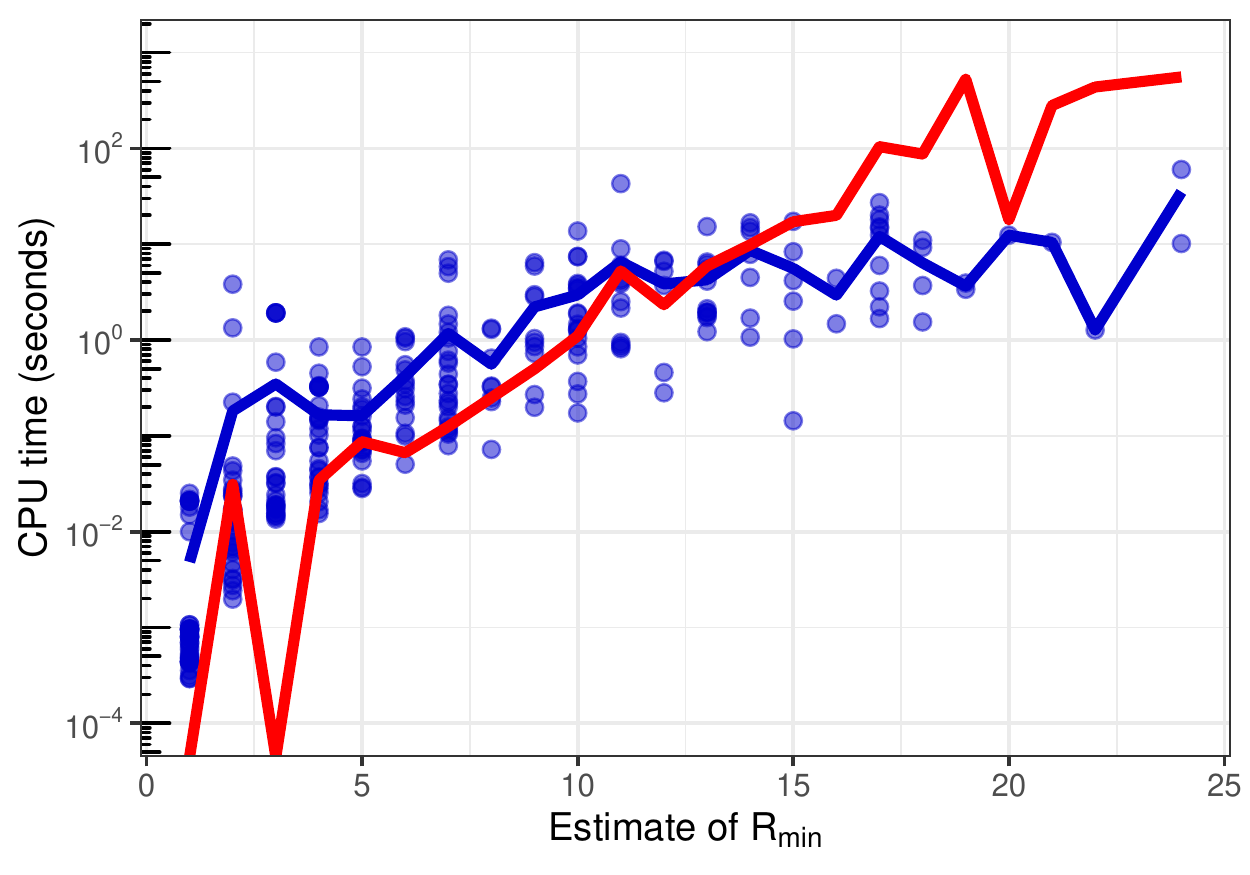}
         \includegraphics[width=0.4\linewidth]{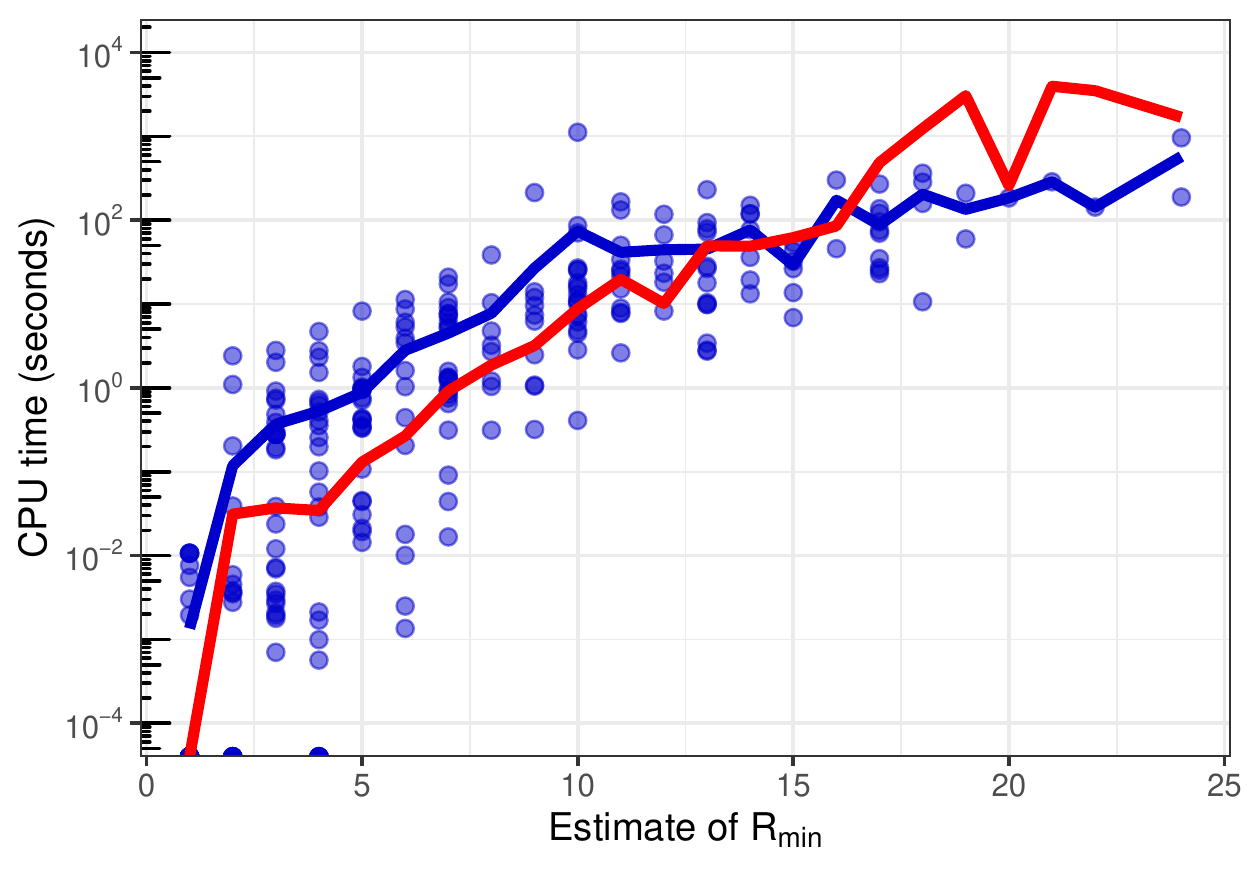}
            \caption{Blue points: time taken to run $Q=20$ iterations of KwARG (left: disallowing recurrent mutations, right: allowing both recombination and recurrent mutation). Blue lines: mean values. Red line: mean run time of SHRUB (left) and SHRUB-GC (right).  Time in seconds is given on a log scale.} \label{time_shrub}
\end{figure}

KwARG obtained solutions at least as good as SHRUB's in 292 (97.3\%) of 300 cases, outperforming it in 35 (11.7\%) instances. KwARG obtained solutions at least as good as SHRUB-GC in 296 (98.7\%) cases, outperforming it in 2 instances. The results and the run times are illustrated in Figures \ref{shrub} and \ref{time_shrub}. On average, for relatively small and simple datasets, KwARG takes approximately the same time per one iteration as a run of SHRUB or SHRUB-GC, and outperforms both programs on more complex datasets.

\section{Comparison to tsinfer, RENT+, and ARGweaver} \label{app_comparisons_s}

Datasets were simulated using msprime under the infinite sites assumption (parameters: $N_e=10\,000$, 20 sequences of length 1\,000bp), with a range of recombination rates ($\{ 1 \cdot 10^{-7}, 2 \cdot 10^{-7}, 4 \cdot 10^{-7}, 8 \cdot 10^{-7}, 1.6 \cdot 10^{-6} \}$ per site per generation) and mutation rates ($\{ 5 \cdot 10^{-8}, 1 \cdot 10^{-7}, 2 \cdot 10^{-7}, 4 \cdot 10^{-7}, 8 \cdot 10^{-7}, 1.6 \cdot 10^{-6}, 3.2 \cdot 10^{-6}, 6.4 \cdot 10^{-6}, 1.28 \cdot 10^{-5} \} $ per site per generation). These parameters were chosen to cover a broad range of the simulated number of recombinations and mutations. 100 datasets were simulated for each combination of rates. 

RENT+, tsinfer, ARGweaver, and KwARG were run on each dataset. For tsinfer, the ancestral state must be specified at each variable site, and was set to the simulated truth. ARGweaver requires the specification of mutation and recombination rates; these were set to the simulation parameters used. ARGweaver was run for 1\,200 iterations, discarding the first 1\,000 as burn-in, and then sampling ARGs with intervals of 20 steps (obtaining 10 in total). KwARG was run for one iteration per dataset, with the parameters $T=30$, $C_{SE} = C_{RM} = \infty$, and the known ancestral sequence set as the root.

\begin{figure}[htbp!]
\centering
        \includegraphics[width=0.27\linewidth]{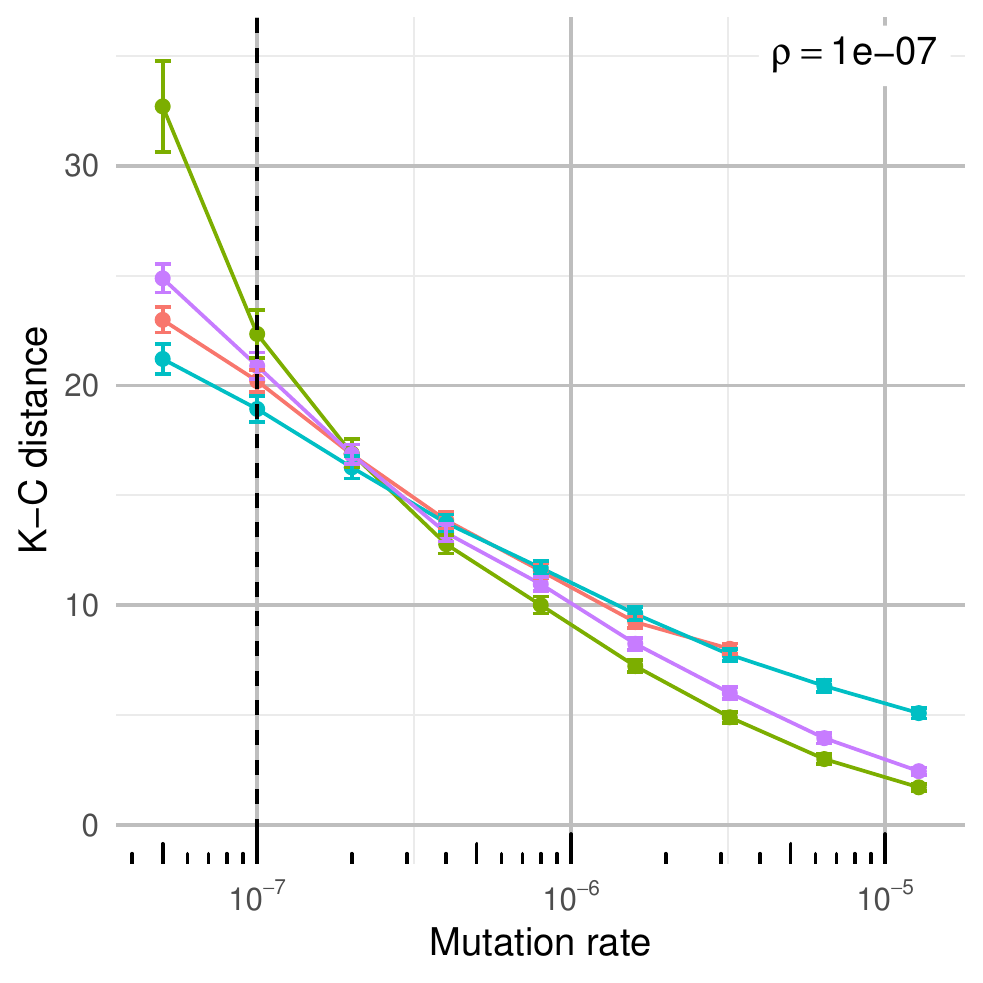}
        \includegraphics[width=0.27\linewidth]{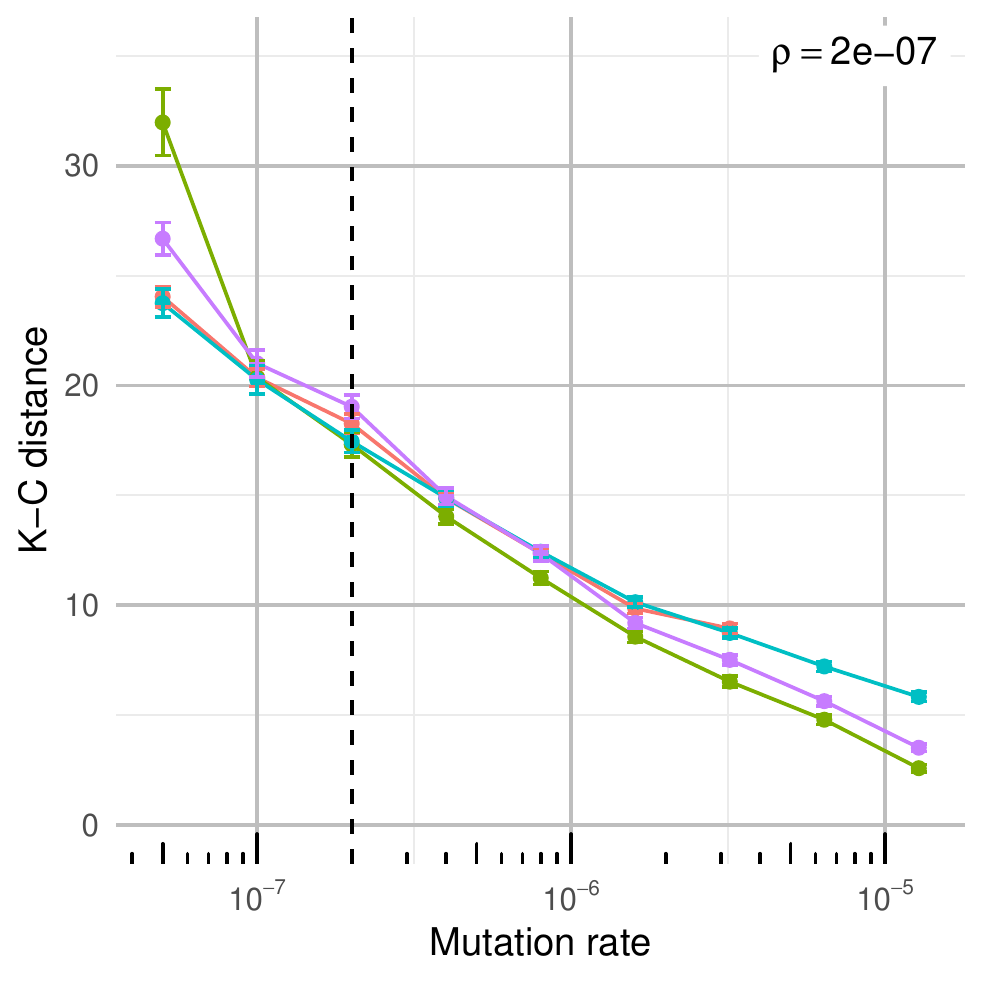}
        \includegraphics[width=0.27\linewidth]{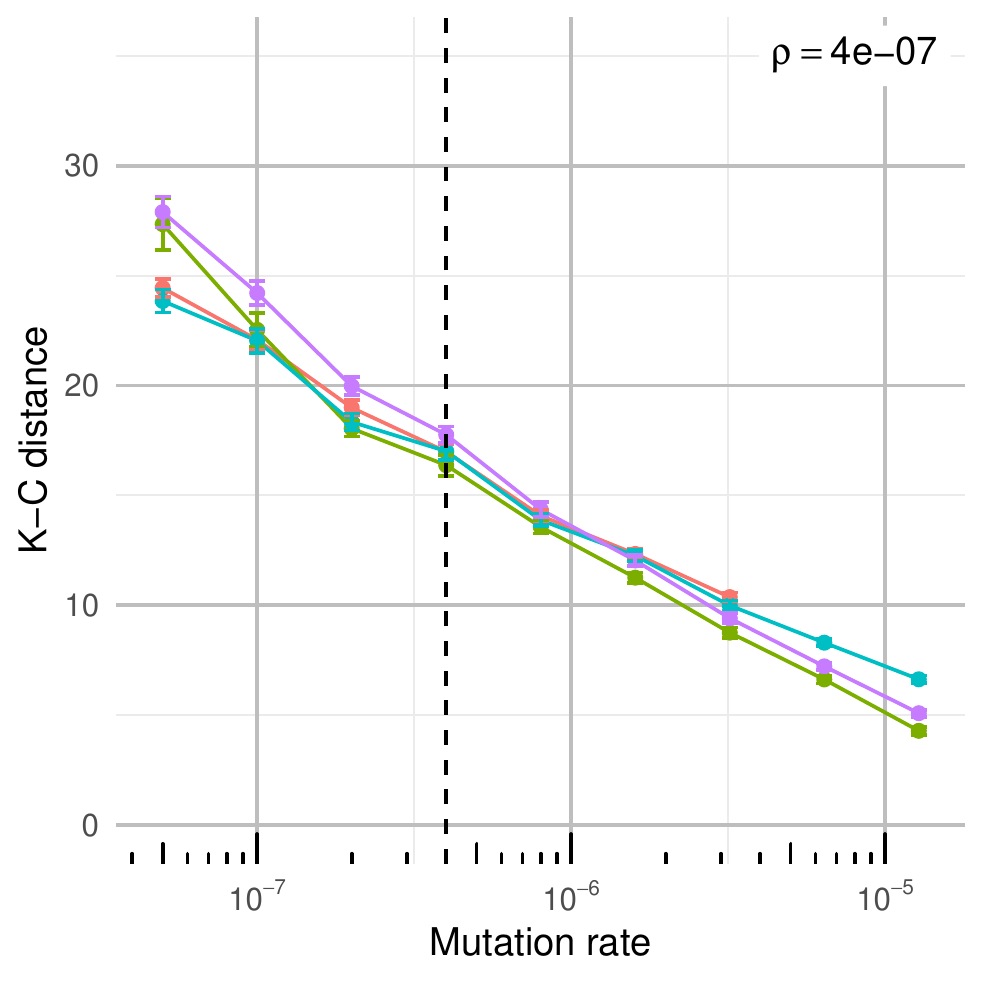}
        \includegraphics[width=0.27\linewidth]{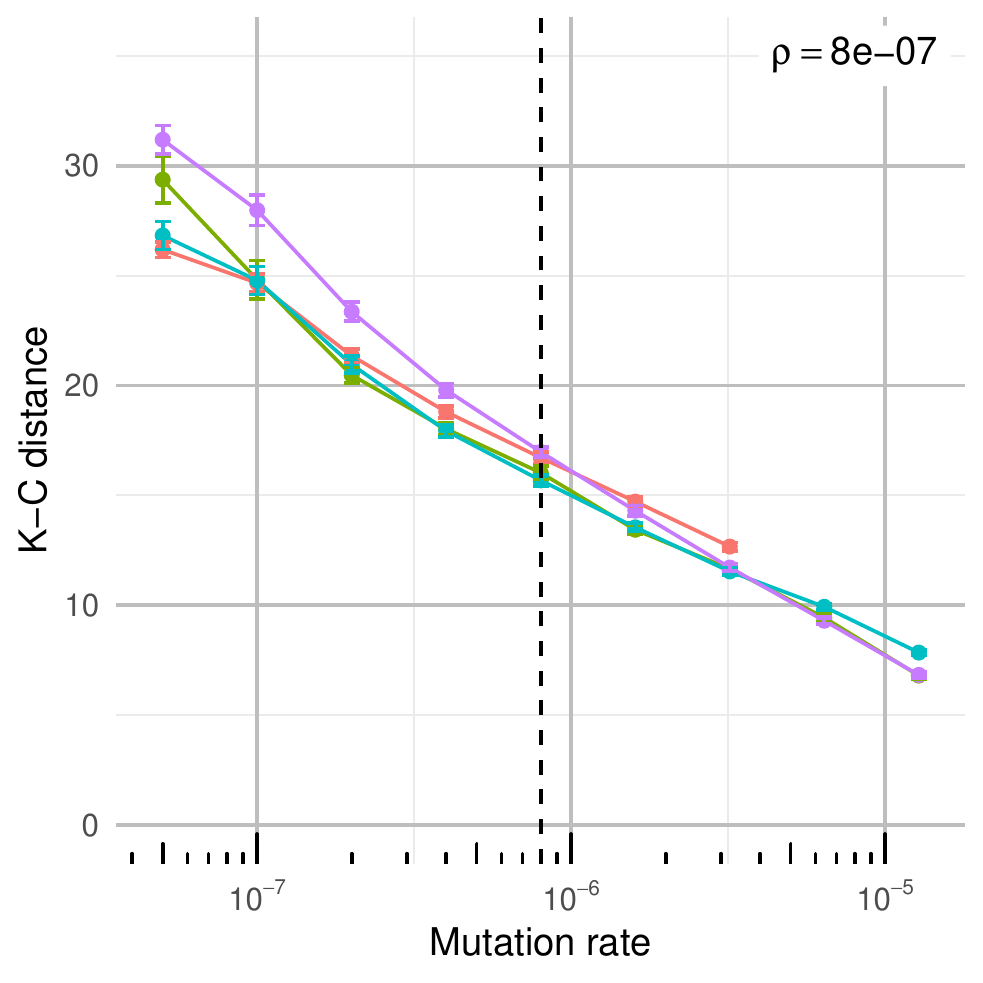}
        \includegraphics[width=0.27\linewidth]{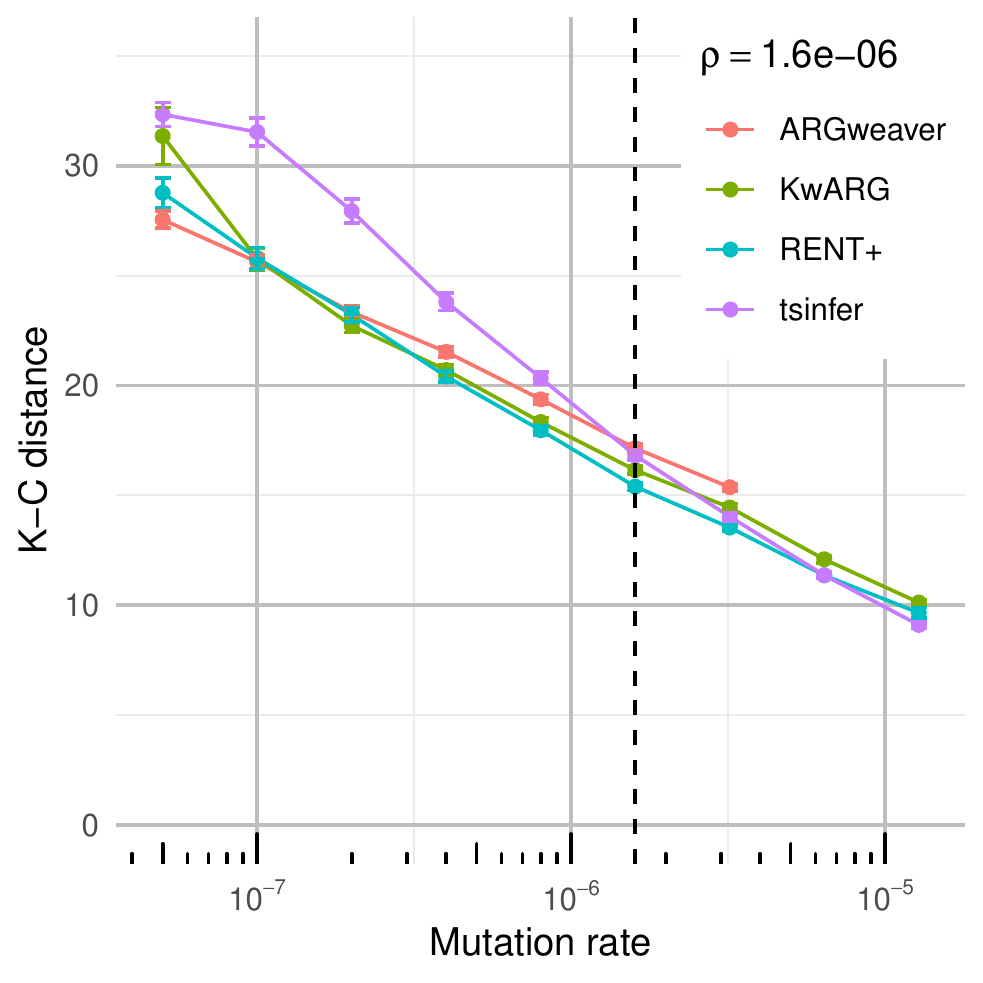}
    \caption{Comparison of performance in local tree recovery. Dashed vertical lines show the value of the recombination rate in each panel. Points correspond to mean values; error bars show mean $\pm$ standard error. ARGweaver results not shown past $\mu = 3.2 \cdot 10^{-6}$ due to prohibitively long run time. Lower K-C distance indicates better accuracy.} \label{app_comparisons}
\end{figure}

\begin{figure}[htbp!]
\centering
        \includegraphics[width=0.27\linewidth]{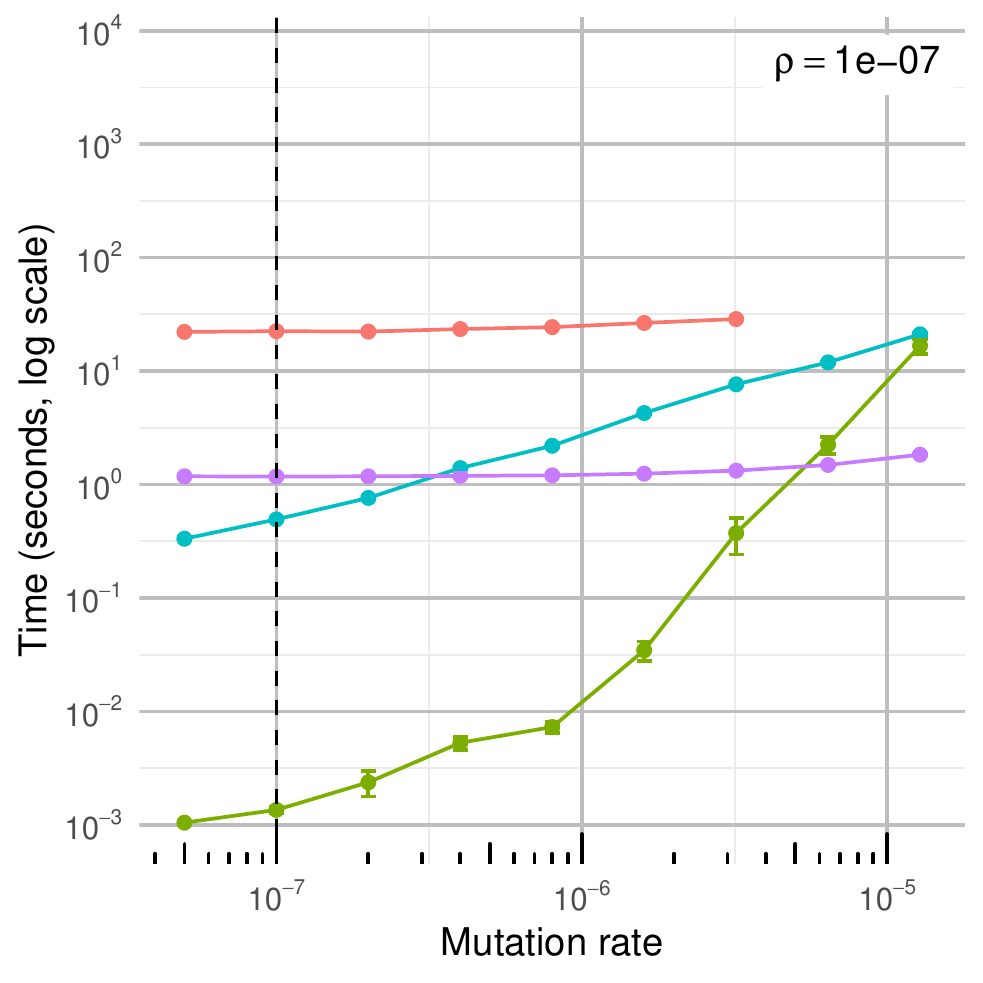}
        \includegraphics[width=0.27\linewidth]{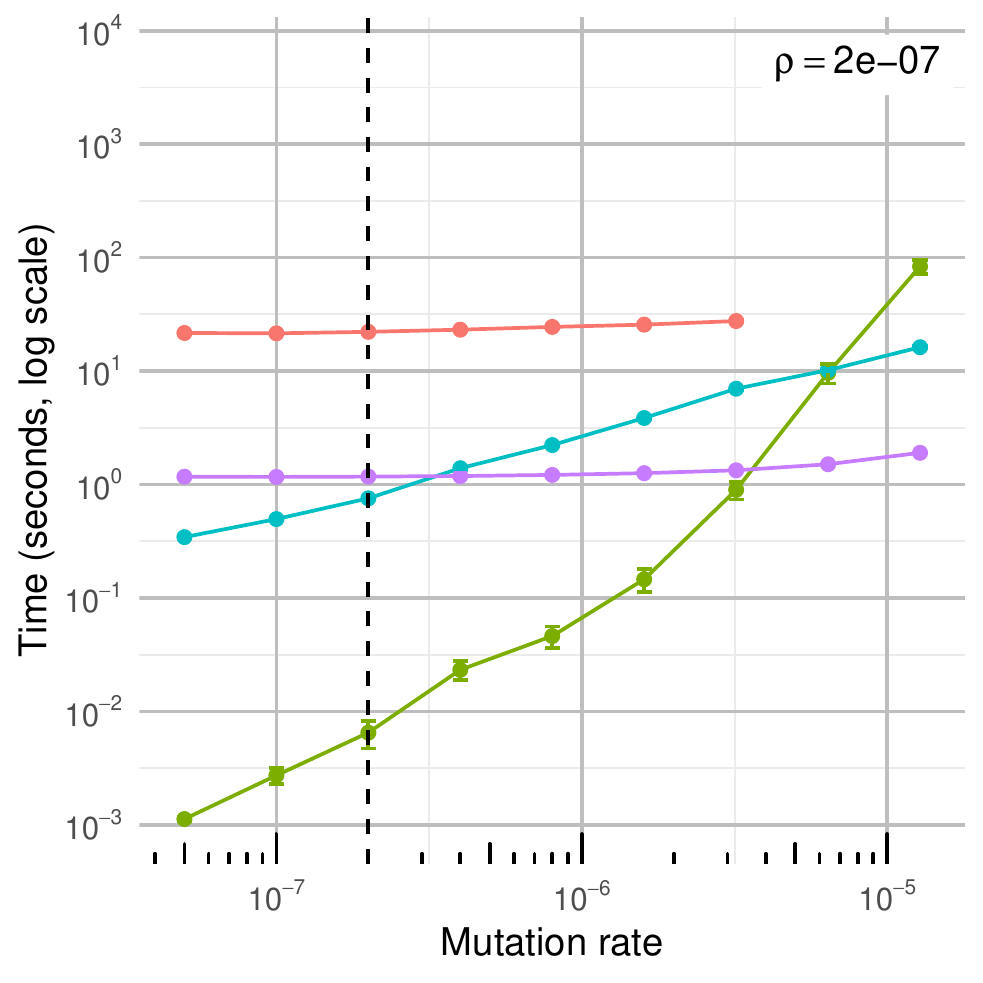}
        \includegraphics[width=0.27\linewidth]{times_4e-07.pdf}
        \includegraphics[width=0.27\linewidth]{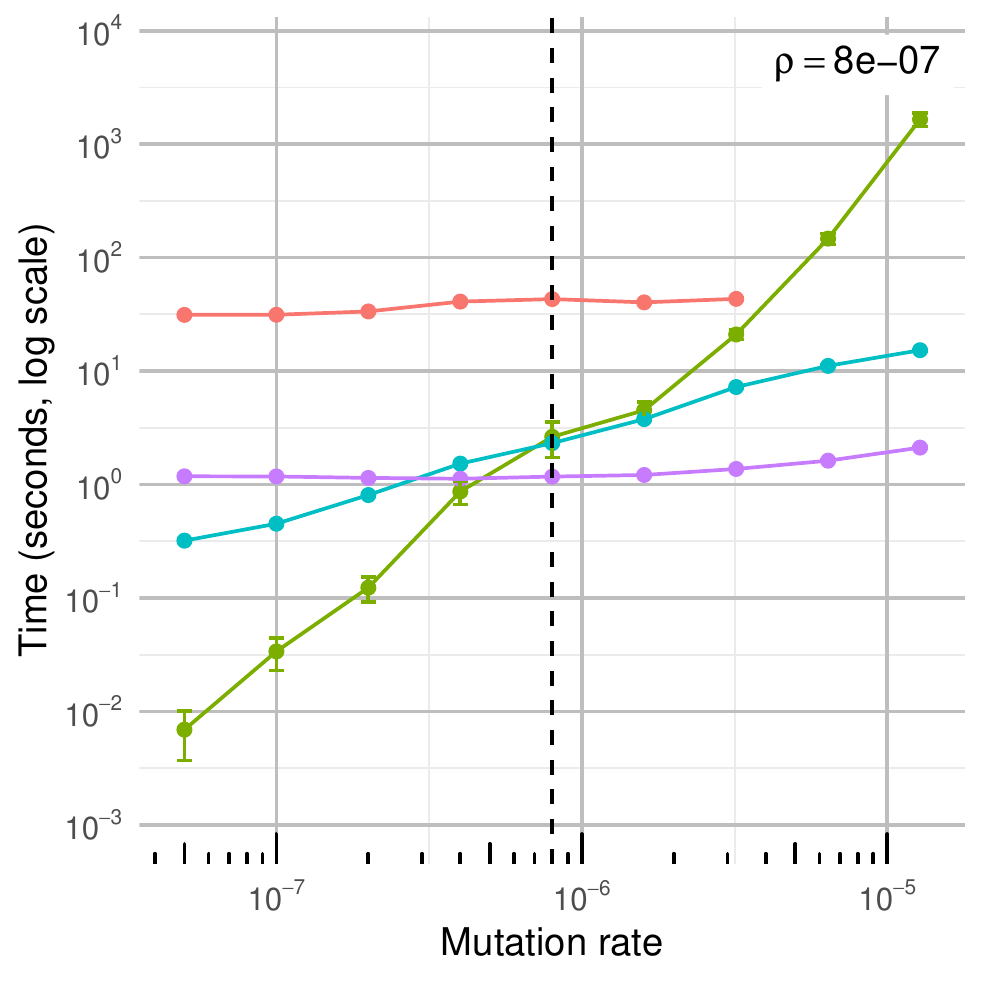}
         \includegraphics[width=0.27\linewidth]{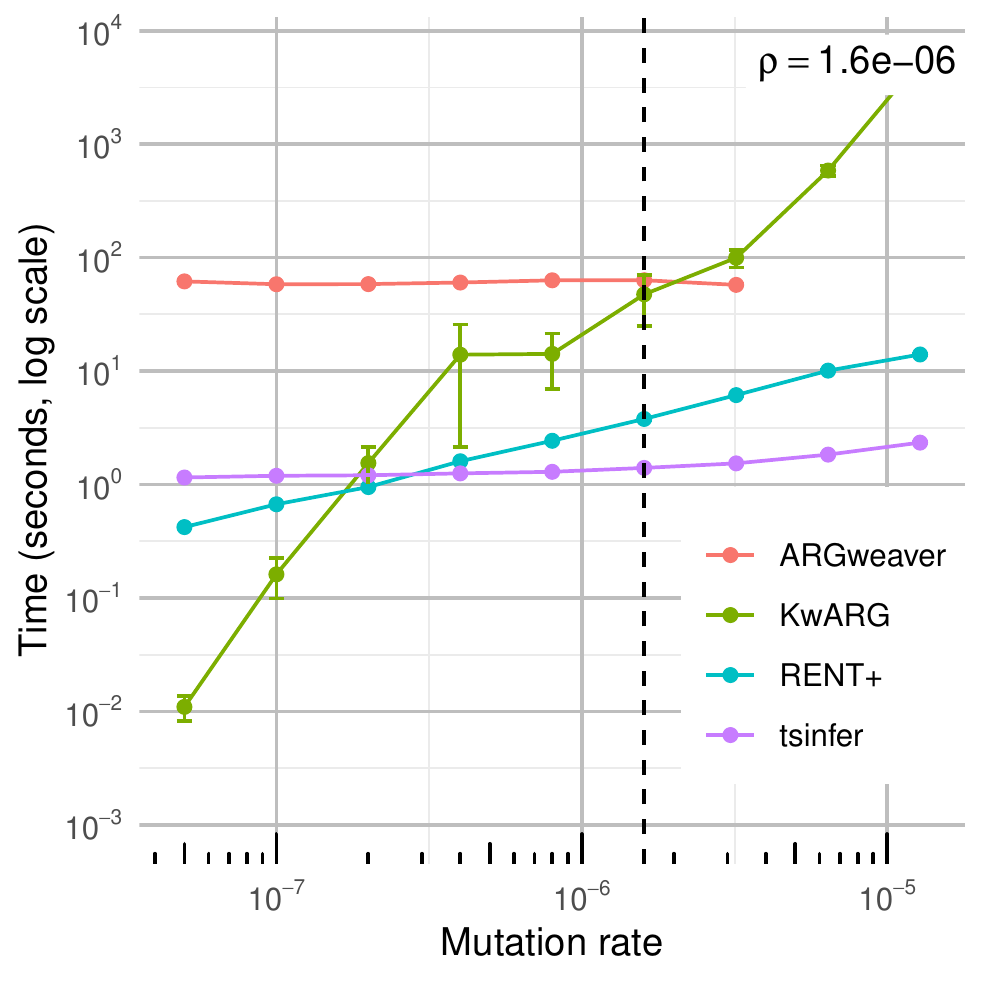}
    \caption{Comparison of time taken per dataset. Points show mean run time averaged over 100 datasets for each combination of rate parameters. Error bars show mean $\pm$ standard error. } \label{app_comparisons_time}
\end{figure}

For each dataset, the local trees output by each program were then compared to the simulated true trees, by calculating the Kendall--Colijn metric at each variable site position. As tsinfer can output trees with unresolved polytomies, these were resolved randomly before calculating the metric for the sake of fair comparison. The mean was then calculated across sites, and for each combination of recombination and mutation rate the metric was averaged across the datasets. The results are presented in Figure \ref{app_comparisons}. A comparison of the run times of the programs used is illustrated in Figure \ref{app_comparisons_time}.

\section{Time complexity} \label{time}

\begin{figure}[htbp!]
\centering
        \includegraphics[width=0.4\linewidth]{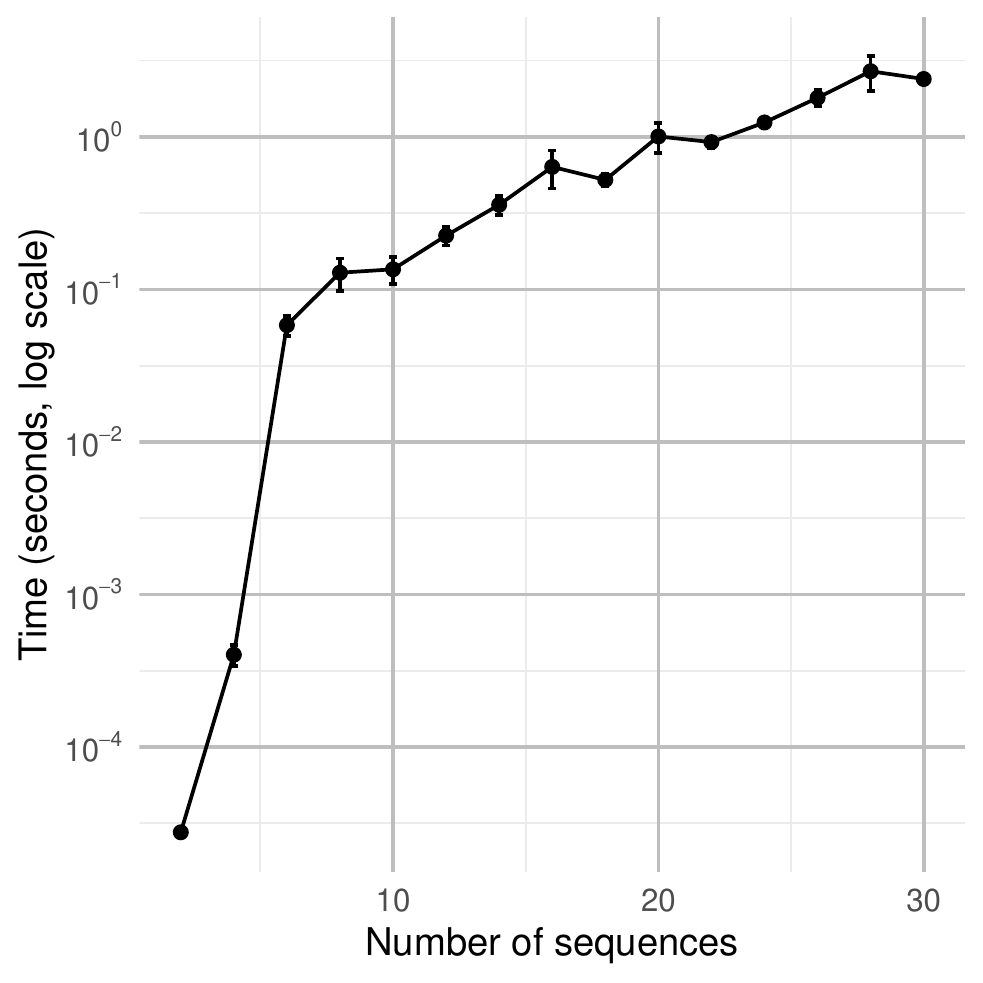}
         \includegraphics[width=0.4\linewidth]{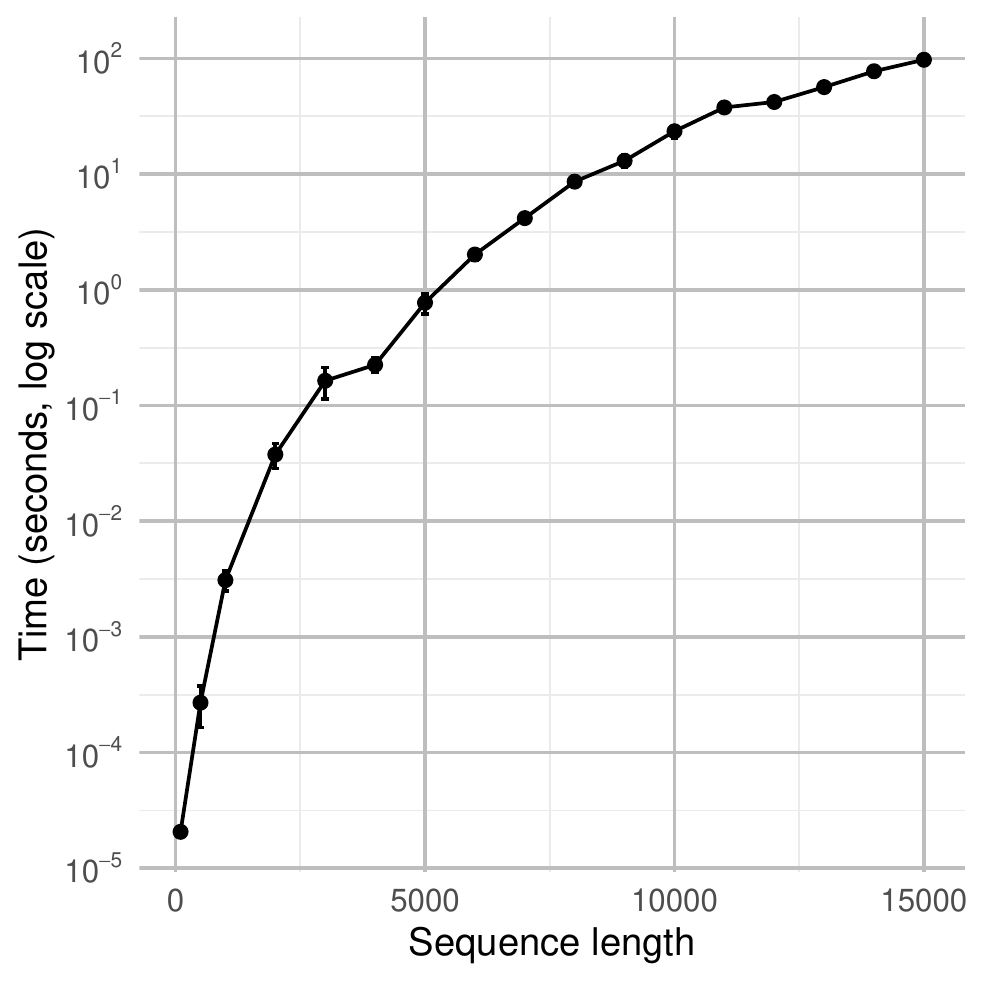}
            \caption{Run time versus number of sequences (left panel) and sequence length (right panel). Lines show mean run time over 500 (100) datasets; error bars show mean $\pm$ standard error.} \label{time_complexity}
\end{figure}

The scaling of KwARG's run time was investigated through simulation. First, we fixed the sequence length at 5\,000bp, and simulated datasets with varying numbers of sequences (from 2 to 30) using msprime, with the infinite sites assumption (parameters: $N_e$ = 10\,000, mutation rate $2 \cdot 10^{-7}$ per site per generation, recombination rate $2 \cdot 10^{-7}$ per site per generation). 500 simulations were carried out for each number of sequences; for each dataset, KwARG was run once and the runtime recorded. The results are presented in the left panel of Figure \ref{time_complexity}. KwARG runs very quickly when the number of sequences is very low, and shows roughly exponential growth in run time when the number of sequences is $6$ or more.

Next, we fixed the number of sequences at 20, and simulated datasets with varying sequence lengths (from 100 to 15\,000bp) using msprime, with the infinite sites assumption (same parameters as above). 100 simulations were carried out for each sequence length; for each dataset, KwARG was run once and the runtime recorded. The results are presented in the right panel of Figure \ref{time_complexity}. After an initial exponential increase (due to small datasets taking very little time per iteration), the run time scales roughly linearly in sequence length.

\end{document}